\documentclass[twocolumn]{aastex631}

\shorttitle{Ly$\alpha$ profiles of $z\simeq2-3$ EELGs}
\shortauthors{Tang et al.}

\begin{document}

\title{Ly$\alpha$ Emission Line Profiles of Extreme [O~{\small III}] Emitting Galaxies at $z\gtrsim2$: Implications for Ly$\alpha$ Visibility in the Reionization Era}

\author{Mengtao Tang} 
\affiliation{Steward Observatory, University of Arizona, 933 N Cherry Ave, Tucson, AZ 85721, USA}
\affiliation{Department of Physics and Astronomy, University College London, Gower Street, London WC1E 6BT, UK}

\author{Daniel P. Stark}
\affiliation{Steward Observatory, University of Arizona, 933 N Cherry Ave, Tucson, AZ 85721, USA}

\author{Richard S. Ellis}
\affiliation{Department of Physics and Astronomy, University College London, Gower Street, London WC1E 6BT, UK}

\author{Michael W. Topping}
\affiliation{Steward Observatory, University of Arizona, 933 N Cherry Ave, Tucson, AZ 85721, USA}

\author{Charlotte Mason}
\affiliation{Cosmic Dawn Center (DAWN), Niels Bohr Institute, University of Copenhagen, Jagtvej 128, 2200 K{\o{}}benhavn N, Denmark}

\author{Zhihui Li}
\affiliation{Cahill Center for Astronomy and Astrophysics, California Institute of Technology, 1200 E California Blvd, MC 249-17, Pasadena, CA 91125, USA}

\author{Ad\`ele Plat}
\affiliation{Institute for Physics, Laboratory for Galaxy Evolution and Spectral Modelling, Ecole Polytechnique Federale de Lausanne, Observatoire de Sauverny, Chemin Pegasi 51, CH-1290 Versoix, Switzerland}



\begin{abstract}
{\it JWST} observations have recently begun delivering the first samples of Ly$\alpha$ velocity profile measurements at $z>6$, opening a new window on the reionization process. Interpretation of $z\gtrsim6$ line profiles is currently stunted by limitations in our knowledge of the intrinsic Ly$\alpha$ profile (before encountering the IGM) of the galaxies that are common at $z\gtrsim6$. To overcome this shortcoming, we have obtained resolved ($R\sim3900$) Ly$\alpha$ spectroscopy of $42$ galaxies at $z=2.1-3.4$ with similar properties as are seen at $z>6$. We quantify a variety of Ly$\alpha$ profile statistics as a function of [O~{\small III}]+H$\beta$ EW. Our spectra reveal a new population of $z\simeq 2-3$ galaxies with large [O~{\small III}]+H$\beta$ EWs ($>1200$~\AA) and a large fraction of Ly$\alpha$ flux emerging near the systemic redshift (peak velocity $\simeq0$~km~s$^{-1}$). These spectra indicate that low density neutral hydrogen channels are able to form in a subset of low mass galaxies ($\lesssim1\times10^8\ M_{\odot}$) that experience a burst of star formation (sSFR $>100$~Gyr$^{-1}$). Other extreme [O~{\small III}] emitters show weaker Ly$\alpha$ that is shifted to higher velocities ($\simeq240$~km~s$^{-1}$) with little emission near line center. We investigate the impact the IGM is likely to have on these intrinsic line profiles in the reionization era, finding that the centrally peaked Ly$\alpha$ emitters should be strongly attenuated at $z\gtrsim5$. We show that these line profiles are particularly sensitive to the impact of resonant scattering from infalling IGM and can be strongly attenuated even when the IGM is highly ionized at $z\simeq 5$. We compare these expectations against a new database of $z\gtrsim6.5$ galaxies with robust velocity profiles measured with {\it JWST}/NIRSpec.
\end{abstract}



\section{Introduction} \label{sec:intro}

Studying the reionization of hydrogen in the intergalactic medium (IGM) provides important clues to understanding the early history of cosmic structure formation. Over the past two decades, numerous observational efforts have been devoted to studying the connection between galaxy formation and cosmic reionization \citep{Stark2016,Robertson2022}. Ly$\alpha$ emission lines from high-redshift galaxies provide a useful tool to probe the neutral hydrogen (H~{\small I}) in the IGM \citep{Dijkstra2014,Ouchi2020}. Because of the strong cross section for scattering with neutral hydrogen, Ly$\alpha$ photons emitted from galaxies at redshifts where the IGM is mostly neutral should be strongly suppressed \citep[e.g.,][]{Miralda-Escude1998,McQuinn2007,Mesinger2015}. 

Spectroscopic observations have revealed that the fraction of galaxies showing prominent Ly$\alpha$ emission (e.g., with equivalent width EW $>25$~\AA) declines significantly from $z\simeq6$ to $z\gtrsim7$, consistent with expectations if the IGM is highly neutral (neutral fraction $x_{\rm HI}\gtrsim0.5$) at $z\gtrsim7$ and becomes highly ionized at $z\simeq6$ \citep[e.g.,][]{Stark2010,Caruana2014,Schenker2014,Pentericci2018,Mason2019,Nakane2023,Jones2024}. Such evolution is also supported by studies of the abundance of narrowband-selected Ly$\alpha$ emitters at $z>5.5$ \citep[e.g.,][]{Ouchi2010,Ota2017,Zheng2017,Konno2018,Itoh2018,Goto2021}. This timeline of reionization is consistent with constraints from measurements of the electron scattering optical depth of the cosmic microwave background \citep{Planck2020} and quasar absorption spectra which suggest that the IGM is substantially neutral at $z\gtrsim7$ \citep[e.g.,][]{Banados2018,Davies2018,Wang2020,Yang2020a,Greig2022} and is significantly ionized at $z\simeq5-6$ (e.g., \citealt{McGreer2015,Yang2020b,Bosman2021,Jin2023,Zhu2023}; see \citealt{Fan2023} for a review).

Over the last decade, attention has begun to focus on using Ly$\alpha$ measurements to trace the local reionization process around galaxies at $z\gtrsim7$. Observations have revealed that many ultraviolet (UV) luminous (M$_{\rm UV}\lesssim-21.5$) galaxies at $z\gtrsim7$ have visible Ly$\alpha$ emission \citep[e.g.,][]{Oesch2015,Zitrin2015,Roberts-Borsani2016,Stark2017,Larson2022,Cooper2023}. It has been suggested that these systems trace overdense regions with a high density of faint neighboring galaxies \citep[e.g.,][]{Tilvi2020,Jung2022,Leonova2022,Whitler2023b,Chen2024}, which are able to power large ionized bubbles \citep[e.g.,][]{Wyithe2005,Dayal2009,Castellano2016,Weinberger2018,Endsley2022a}. In this case, Ly$\alpha$ photons will be significantly redshifted before encountering the neutral IGM, boosting the transmission of the line \citep[e.g.,][]{Mesinger2004,Mason2020,Qin2022,Smith2022,Napolitano2024}. There is also evidence that the Ly$\alpha$ peak of these galaxies is offset to a high velocity from the systemic redshift, shifting the Ly$\alpha$ photons far into the damping wing before encountering the neutral IGM \citep[e.g.,][]{Stark2017,Tang2023}. This further boosts the transmission of Ly$\alpha$, countering the attenuation provided by the neutral IGM \citep[e.g.,][]{Mason2018b,Endsley2022b}. Efforts are underway to link Ly$\alpha$ emission in these systems to bubbles sizes \citep[e.g.,][]{Hayes2023,Lu2024}, but such estimates rely on knowledge of how much Ly$\alpha$ is redshifted relative to the galaxy systemic redshift.

Spectroscopy with {\it JWST} \citep{Gardner2023} NIRSpec \citep{Jakobsen2022} has recently pushed the Ly$\alpha$ frontier beyond $z\simeq10$ \citep{Bunker2023a}, while also delivering the first large samples of Ly$\alpha$ profile measurements at $z\gtrsim7$ \citep[e.g.,][]{Bunker2023a,Tang2023,Saxena2024}. Some $z\gtrsim7$ galaxies have been detected with Ly$\alpha$ equivalent widths (EWs) $\simeq10-20$~\AA\ and relatively large Ly$\alpha$ peak velocity offsets ($\gtrsim400$~km~s$^{-1}$; e.g., \citealt{Bunker2023a,Tang2023}), similar to the luminous galaxies studied prior to {\it JWST} \citep[e.g.,][]{Stark2017,Endsley2022b}. However {\it JWST} has also revealed discovery of a new class of systems at $z\gtrsim 7$ \citep{Saxena2023,Chen2024}, with extremely strong Ly$\alpha$ emission (EW $\simeq300-400$~\AA) which may be escaping with low Ly$\alpha$ velocity offsets ($\simeq100$~km~s$^{-1}$). If such strong Ly$\alpha$ is observed near the systemic redshift, it would require that the emitting galaxy resides in a large ionized region ($R\gtrsim3$~pMpc; \citealt{Saxena2023}), allowing the line profile to escape with minimal processing by the IGM. 

Reliably linking Ly$\alpha$ velocity offsets to constraints on bubble sizes relies on knowledge of the full range of factors modulating the Ly$\alpha$ profiles in reionization-era galaxies. Before Ly$\alpha$ photons encounter the IGM, the H~{\small I} distribution in the ISM and the CGM resonantly scatters the Ly$\alpha$ photons emitted from H~{\small II} regions. The profiles we are now observing at $z\gtrsim7$ will have been further altered by scattering from the neutral IGM. Even at $z\simeq5$ when the IGM is highly ionized, the residual H~{\small I} in the IGM will attenuate the Ly$\alpha$ emission near line center via resonant scattering \citep[e.g.,][]{Gunn1965}. Without a detailed understanding of the range of intrinsic\footnote{In this paper, we define the intrinsic Ly$\alpha$ profile as that which emerges from the ISM and CGM of the host galaxy prior to interaction with the IGM.} Ly$\alpha$ spectral shapes in galaxies typical of the reionization-era, it is difficult to reliably assess the impact of the IGM on the observed Ly$\alpha$ profiles at $z\gtrsim6$, stunting efforts to infer ionized bubble sizes around known Ly$\alpha$ emitters.

High-resolution ($R\gtrsim4000$) Ly$\alpha$ spectroscopy of galaxies at lower redshifts ($z\simeq2-3$) provides our best path toward understanding the range of intrinsic Ly$\alpha$ profiles that are likely present in reionization-era galaxies. While such spectra have been obtained for typical galaxies at $z\simeq2-3$ \citep[e.g.,][]{Shapley2003,Steidel2010,Erb2014,Trainor2015,Matthee2021}, they do not exist for galaxies with properties similar to that seen at $z\gtrsim6$. In this paper, we seek to build such a Ly$\alpha$ spectral library at $z\simeq2-3$. A key feature of reionization-era galaxies is intense [O~{\small III}]+H$\beta$ line emission (with median rest-frame EW $z\simeq700-800$~\AA; e.g., \citealt{Labbe2013,DeBarros2019,Endsley2021a,Endsley2023}), as expected in moderately metal poor systems with young stellar populations. \citet{Du2020} and \citet{Tang2021b} have presented a first step toward studying the Ly$\alpha$ emission of this population, with medium-resolution ($R\simeq1000$) Ly$\alpha$ spectroscopy of $z\sim2-3$ extreme emission line galaxies (EELGs) spanning the full range of [O~{\small III}]+H$\beta$ EWs expected at $z\simeq7-8$ (EW$_{{\rm [OIII]+H}\beta}\simeq300-3000$~\AA; e.g., \citealt{Endsley2021a,Endsley2023}). In this work, we present high-resolution ($R\simeq4000$) spectroscopy of $42$ EELGs at $z=2.1-3.4$, enabling characterization of resolved line profiles. Using this dataset, we explore the range of Ly$\alpha$ profiles seen in galaxies with different [O~{\small III}]+H$\beta$ EWs. The dataset allows insight into the intrinsic Ly$\alpha$ profiles (and hence the H~{\small I} distribution) that are likely in reionization-era galaxies. We use our spectral library to discuss how the $z\gtrsim5$ IGM is likely to alter these line profiles. We compare these expectations against the existing sample of $z\gtrsim 6.5$ Ly$\alpha$ emitters with robust velocity profiles from {\it JWST}/NIRSpec grating spectroscopy. 

The organization of this paper is as follows. In Section~\ref{sec:OA}, we describe the observations and the resolved Ly$\alpha$ spectroscopy of $z=2.1-3.4$ EELGs. We present the Ly$\alpha$ profiles of sources in our sample and discuss the constraints on the H~{\small I} distribution in Section~\ref{sec:result}. We then discuss the implications for the Ly$\alpha$ profiles of $z\gtrsim7$ galaxies and the Ly$\alpha$ visibility in the reionization era in Section~\ref{sec:discussion}. Finally, we summarize our conclusions in Section~\ref{sec:summary}. We adopt a $\Lambda$-dominated, flat universe with $\Omega_{\Lambda}=0.7$, $\Omega_{\rm M}=0.3$, and $H_0=70$~km~s$^{-1}$~Mpc$^{-1}$. All magnitudes in this paper are quoted in the AB system \citet{Oke1983}, and all EWs are quoted in the rest frame.


\section{Observations and Analysis} \label{sec:OA}

In this work, we aim to characterize the Ly$\alpha$ emission line profiles of low mass galaxies with extreme [O~{\small III}] emission lines at $z\simeq2-3$ using high-resolution ($R\simeq4000$) spectroscopy. We describe the sample selection and spectroscopic observations in Section~\ref{sec:spectra}, and then present the resolved Ly$\alpha$ profiles in Section~\ref{sec:measurement}.


\begin{figure*}
\includegraphics[width=\linewidth]{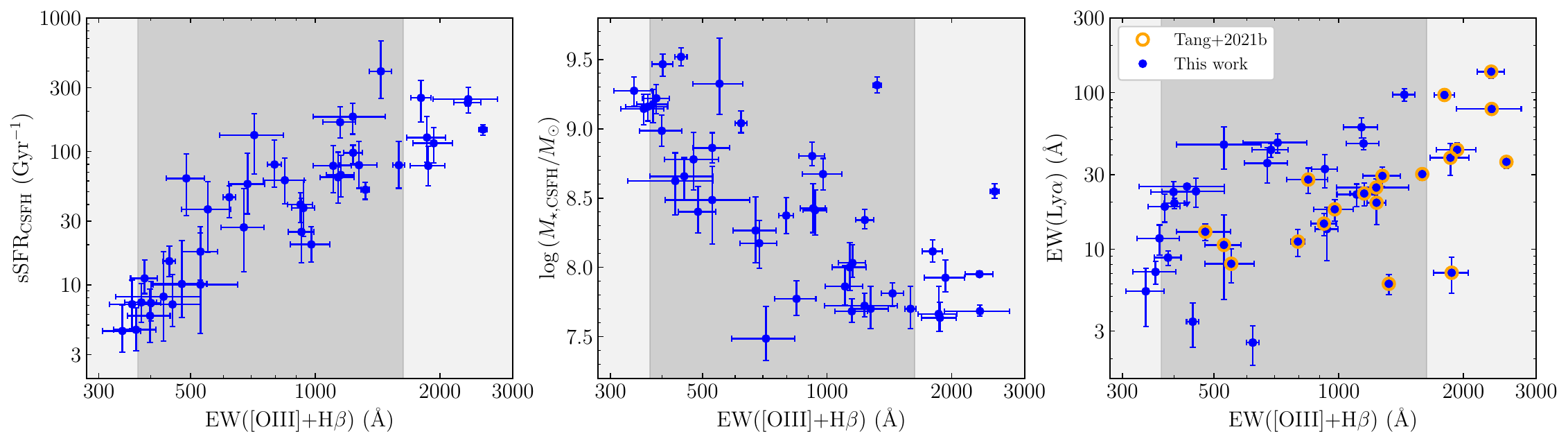}
\caption{Specific star formation rate (sSFR; left panel), stellar mass (middle panel), and Ly$\alpha$ EW (right panel) as a function of [O~{\scriptsize III}]+H$\beta$ EW for galaxies in our resolved Ly$\alpha$ spectroscopic sample at $z=2.1-3.4$. Stellar masses and sSFRs are derived from the {\tt BEAGLE} models assuming CSFH. In the right panel, objects that have already obtained low-resolution Ly$\alpha$ spectra in \citet{Tang2021b} are marked by orange open circles. We mark the [O~{\scriptsize III}]+H$\beta$ EW range that is typical at $z>6$ with the dark grey ($68\%$ within the median value) and the light grey shaded regions ($95\%$ within the median) based on the [O~{\scriptsize III}]+H$\beta$ EW distribution at $z>6$ presented in \citet{Endsley2023}. Galaxies with larger [O~{\scriptsize III}]+H$\beta$ EWs tend to have larger sSFRs, lower stellar masses, and stronger Ly$\alpha$ emission.}
\label{fig:ssfr_mass_lyaew_o3hbew}
\end{figure*}

\subsection{Spectroscopic Observations} \label{sec:spectra}

The Ly$\alpha$ spectra studied in this paper follow a large rest-frame optical spectroscopic survey of EELGs at $z=1.3-3.7$ \citep{Tang2019,Tang2022} in the Cosmic Assembly Near-infrared Deep Extragalactic Legacy Survey (CANDELS; \citealt{Grogin2011,Koekemoer2011}) fields. The sample of EELGs was identified based on the [O~{\small III}] EWs inferred from {\it HST} grism spectra (at $z=1.3-2.4$) or the $K$-band flux excess (at $z=3.1-3.7$). EELGs are required to have large rest-frame [O~{\small III}]~$\lambda\lambda4959,5007$ EWs with values $\simeq300-3000$~\AA, which are chosen to match the range expected to be common in reionization-era galaxies \citep[e.g.,][]{Endsley2023}. Sources that harbour active galactic nuclei (AGN) were removed based on their X-ray detections by matching the coordinates to the {\it Chandra} X-ray catalogs. We direct the reader to \citet{Tang2019} for the full description of the EELG sample selection. A low-resolution ($R\simeq1000$) rest-frame UV spectroscopic study of the EELGs has been presented in \citet{Tang2021a,Tang2021b}. Here we measure the Ly$\alpha$ line profiles of EELGs using high-resolution spectroscopy. 

The resolved Ly$\alpha$ spectra of EELGs were taken from the Binospec \citep{Fabricant2019} on the MMT telescope with multi-slit spectroscopy mode. We utilized the $1000$ lines mm$^{-1}$ grism blazed at $13.75^{\circ}$ with a wavelength coverage from $3700$ to $5400$~\AA\ (centered at $4500$~\AA). This wavelength range allows us to measure Ly$\alpha$ emission line at $z=2.1-3.4$. We designed one multi-slit mask in the Ultra Deep Survey (UDS) field (centered at R.A. $=$ 02:17:26.8 and Decl. $=-$05:18:52.0 with position angle PA $=-78^{\circ}$), targeting $44$ EELGs at $z=2.1-3.4$ to measure their Ly$\alpha$ emission. The {\it HST} $i_{\rm F814W}$ magnitudes of these $44$ targets range from $23.9$ to $26.9$ AB mag with a median of $i_{\rm F814W}=25.5$, corresponding to absolute UV magnitudes M$_{\rm UV}=-21.7$ to $-18.1$ (median M$_{\rm UV}=-19.5$). We also filled the mask with $16$ EELGs at lower redshift ($z=1.4-2$) to measure C~{\small III}] emission, thereby continuing our ongoing survey targeting UV metal emission lines in EELGs \citep{Tang2021a}. The targets were placed on the mask using the selection function introduced in \citet{Tang2019,Tang2021a}. The target priority was adjusted based on their [O~{\small III}] EWs, and those with the largest EWs ([O~{\small III}]+H$\beta$ EW $>1500$~\AA) were given the highest priority as they are very rare and have rest-frame optical spectral properties similar to the luminous Ly$\alpha$ emitters at $z>7$ \citep[e.g.,][]{Roberts-Borsani2016,Stark2017,Tang2023}. We observed this mask in between September and November 2021 with a total on-target integration time of $16$ hours during an average seeing of $1.0$~arcsec. The slit width was set to $1.0$~arcsec, resulting in a spectral resolution of $R=3900$ (corresponding to $\sigma_{\rm instrument}=33$~km~s$^{-1}$), which allows us to resolve the multi-peak nature of Ly$\alpha$ emission line. 

We reduced the Binospec spectra using the publicly available data reduction pipeline\footnote{\url{https://bitbucket.org/chil_sai/binospec}} \citep{Kansky2019}. The pipeline performs flat-fielding, wavelength calibration, sky subtraction, and then the 2D spectra extraction. The 1D spectra extraction and flux calibration were performed following the procedures described in \citet{Tang2021a}. We created 1D spectra from the reduced 2D spectra using a boxcar extraction. We observed spectrophotometric standard stars and the instrumental response was corrected using the sensitivity curve derived observations of standard stars. Slit loss correction was performed using the in-slit light fraction computed from {\it HST} image following the procedures described in \citet{Kriek2015}. We then performed the absolute flux calibration using observations of slit stars, by comparing the slit-loss corrected count rates of slit star spectra with the broadband flux in the \citet{Skelton2014} catalogues. 

Our goal is to measure the Ly$\alpha$ line profiles in EELGs at $z=2.1-3.4$ which can be used as analogs of reionization-era systems. The $44$ targets in our sample span a wide range of [O~{\small III}]+H$\beta$ EW ($=342-2541$~\AA), typical of values expected at $z>6$ \citep[e.g.,][]{Endsley2023}. We derive the stellar population properties (stellar mass, stellar age, and specific star formation rate, sSFR) of the $44$ targets by fitting the broadband photometry from the \citet{Skelton2014} catalogs and available emission line fluxes using the {\tt BEAGLE} tool \citep{Chevallard2016} assuming constant star formation history (CSFH; see \citealt{Tang2019,Tang2021a} for details of modelling procedures). {\tt BEAGLE} uses the latest version of \citet{Bruzual2003} stellar population models and combines with {\tt Cloudy} \citet{Ferland2013} to compute the nebular emission following the methods in \citet{Gutkin2016}. The intense rest-frame optical emission of our targets indicates young ages ($\simeq2-200$~Myr) and large sSFRs ($\simeq4-400$~Gyr$^{-1}$). The stellar mass of our sample spans from $10^{7.5}\ M_{\odot}$ to $10^{9.5}\ M_{\odot}$. Galaxies with larger [O~{\small III}] EWs tend to be lower mass (assuming CSFH) systems with larger sSFRs (left and middle panels of Figure~\ref{fig:ssfr_mass_lyaew_o3hbew}). The median [O~{\small III}]+H$\beta$ EW and sSFR of our sample are $883$~\AA\ and $54$~Gyr$^{-1}$, which are much larger than the average values of typical $z\sim2-3$ galaxies \citep[e.g.,][]{Reddy2012,Marmol-Queralto2016,Santini2017,Boyett2022} but more comparable to typical galaxies at $z\sim7-8$ (\citealt{Labbe2013,DeBarros2019,Endsley2021a,Endsley2023}). In particular, our sample includes $13$ galaxies with very intense optical line emission (EW$_{{\rm [OIII]+H}\beta}>1200$~\AA), a population that is extremely rare at $z\sim2-3$ \citep{Boyett2022} but becomes more common in the reionization era \citep{Smit2015,Endsley2021a,Endsley2023,Bouwens2023}. The light of such galaxies is dominated by very young stellar populations ($\lesssim10$~Myr assuming CSFH, though older stellar populations could be outshined by young stars; e.g., \citealt{Tang2022,Whitler2023a}), as expected for systems that have recently experienced extreme bursts of star formation. 


\begin{figure*}
\includegraphics[width=\linewidth]{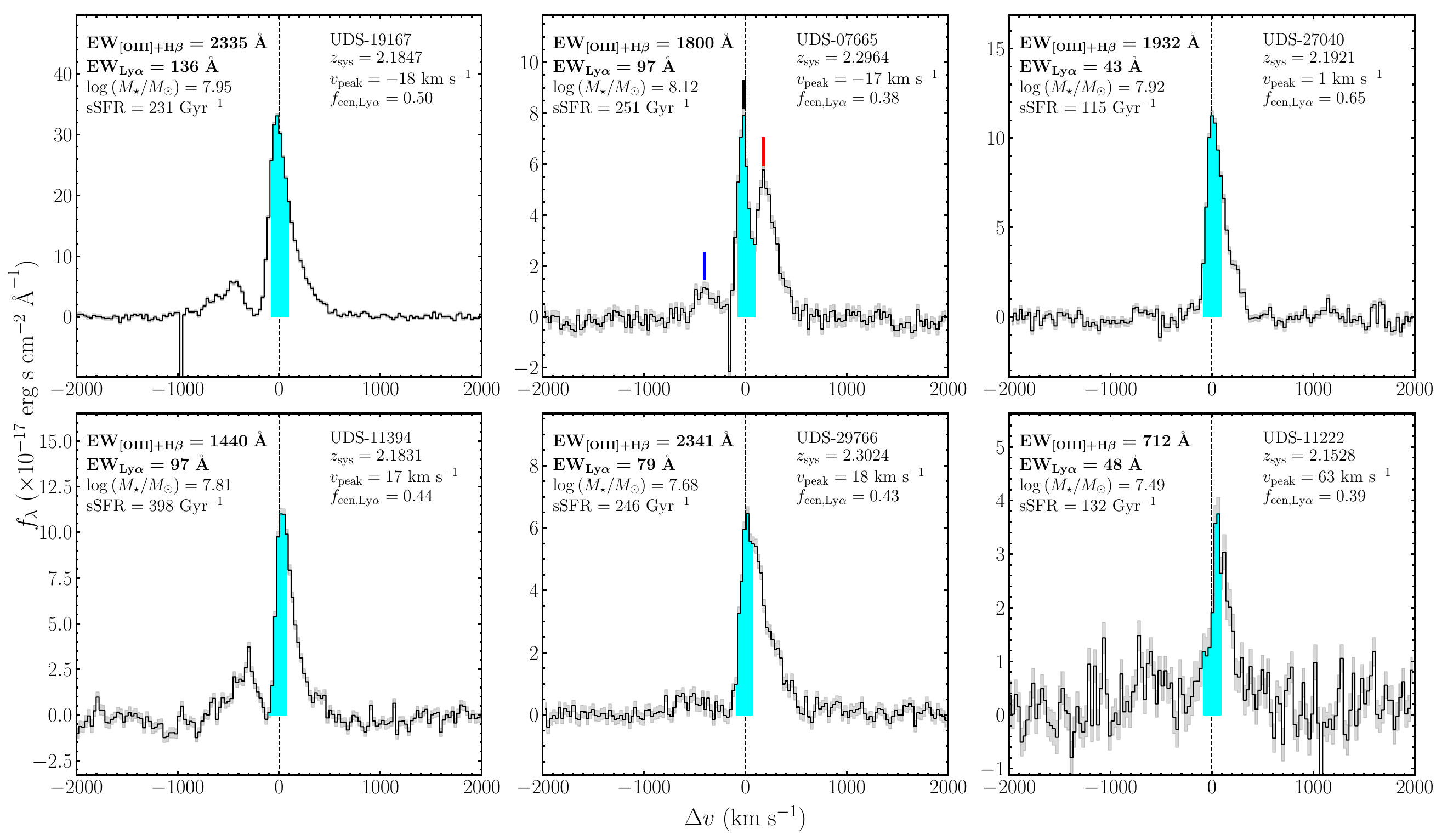}
\caption{Ly$\alpha$ emission profiles of the six EELGs with high Ly$\alpha$ central escape fractions ($f_{{\rm cen,Ly}\alpha}\gtrsim0.4$) and strong Ly$\alpha$ emission (EW$_{{\rm Ly}\alpha}>40$~\AA) at our $z=2.1-3.4$ sample. Cyan shaded regions marking the Ly$\alpha$ photons emitting within $\pm100$~km~s$^{-1}$ of the systemic redshift (black dashed line). Their Ly$\alpha$ peaks are close to the systemic redshifts ($v_{\rm peak}<100$~km~s$^{-1}$), potentially indicating ionized channels in the ISM and the CGM that allow Ly$\alpha$ to escape directly into the IGM. UDS-07665 has a triple-peak Ly$\alpha$ profile, with the blue, central, and red peaks marked by the blue, black, and red lines.}
\label{fig:lya_spec1}
\end{figure*}

We also derive the hydrogen ionizing photon production efficiency ($\xi_{\rm ion}$) of our targets from {\tt BEAGLE} models. Here we use $\xi_{\rm ion}$ defined as the hydrogen ionizing photon production rate ($\dot{N}_{\rm ion}$) per dust-corrected luminosity at rest-frame $1500$~\AA\ ($L_{\rm UV}$, including nebular and stellar continuum; see \citealt{Chevallard2018,Tang2019} for various definition of $\xi_{\rm ion}$). The $\xi_{\rm ion}$ of the $44$ targets ranges from $10^{25.3}$ to $10^{25.9}$~erg$^{-1}$~Hz. For the subset with H$\beta$ and H$\alpha$ emission line measurement, we compare their $\xi_{\rm ion}$ derived from dust-corrected H$\alpha$ luminosity plus $L_{\rm UV}$ \citep{Tang2019} and from {\tt BEAGLE} models. We find both values are consistent. The $\xi_{\rm ion}$ of our sample is higher than the $\xi_{\rm ion}$ of typical star-forming galaxies at $z\sim2$ \citep[e.g.,][]{Matthee2017,Shivaei2018} but comparable to $z>6$ sources \citep[e.g.,][]{Stark2017,Endsley2021a,Stefanon2022,Ning2023,Simmonds2023,Tang2023}, indicating that our EELGs have intense ionizing spectra similar to that seen in the reionization era. 


\begin{figure*}
\includegraphics[width=\linewidth]{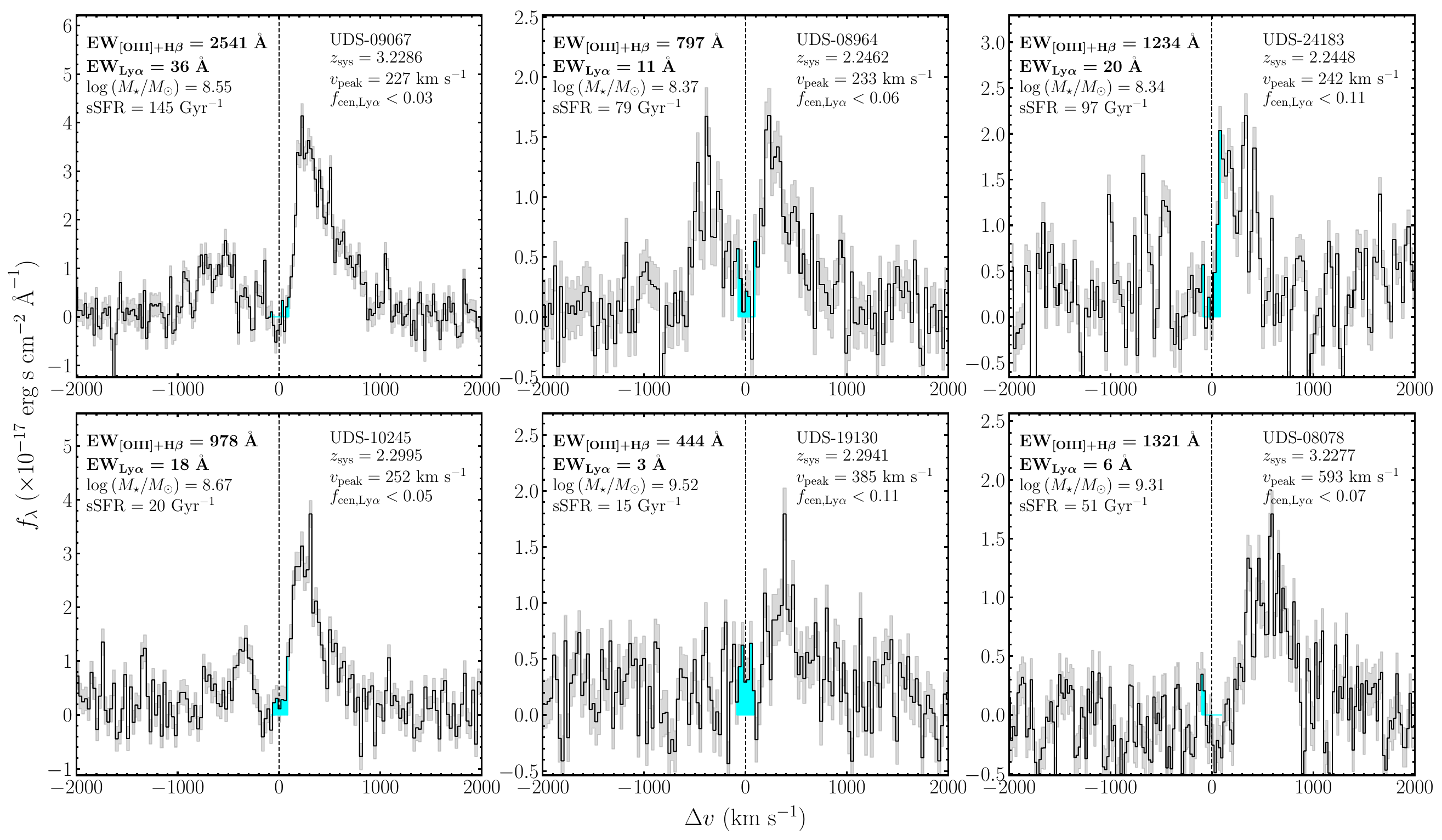}
\caption{Examples of Ly$\alpha$ emission profiles of our $z=2.1-3.4$ EELGs with moderate Ly$\alpha$ emission (EW$_{{\rm Ly}\alpha}=3-40$~\AA) which emit negligible Ly$\alpha$ flux near (within $\pm100$~km~s$^{-1}$) the line center (cyan shaded regions). Spectra are plotted in a similar way as in Figure~\ref{fig:lya_spec1}. Their Ly$\alpha$ peaks are shifted to relatively large velocities ($\gtrsim200$~km~s$^{-1}$), indicating that the H~{\small II} regions are likely covered by denser H~{\small I} gas.}
\label{fig:lya_spec2}
\end{figure*}

We identify Ly$\alpha$ emission lines and compute the Ly$\alpha$ fluxes and EWs by applying the procedures described in \citet{Tang2021b}. Using the redshifts derived by fitting [O~{\small III}]~$\lambda5007$ emission lines from the ground-based (26 galaxies; \citealt{Tang2019}) or {\it HST} grism-based (18 galaxies; \citealt{Momcheva2016}) rest-frame optical spectra, we visually inspect the expected positions of Ly$\alpha$ in the 2D Binospec spectra. Out of the total $44$ EELGs at $z=2.1-3.4$ on the mask, we have detected Ly$\alpha$ emission with S/N $>3$ in $42$ sources. For the remaining $2$ sources lacking Ly$\alpha$ detections, we estimate $3\sigma$ upper limits for the Ly$\alpha$ flux and EW.

The Ly$\alpha$ fluxes are measured from the 1D spectra (examples shown in Figure~\ref{fig:lya_spec1} \& Figure~\ref{fig:lya_spec2}). Due to the complex profile of resolved Ly$\alpha$ emission, we compute the line fluxes for the $42$ galaxies with Ly$\alpha$ detections by directly integrating the flux between rest-frame $1212$~\AA\ and $1220$~\AA. This wavelength window captures the total Ly$\alpha$ flux for all Ly$\alpha$ emitting sources \citep[e.g.,][]{Du2020,Matthee2021}. The measured Ly$\alpha$ fluxes are from $2.4\times10^{-18}$ to $4.2\times10^{-16}$~erg~s$^{-1}$~cm$^{-2}$. For the remaining $2$ galaxies without Ly$\alpha$ detection, the $3\sigma$ upper limit of Ly$\alpha$ flux is derived by integrating the error spectrum in quadrature over rest-frame $1199.9-1228.8$~\AA\ \citep{Kornei2010}. The $3\sigma$ Ly$\alpha$ flux limits of these 2 objects are $2.0\times10^{-17}$ and $2.9\times10^{-17}$~erg~s$^{-1}$~cm$^{-2}$. 

Using the measured Ly$\alpha$ fluxes, we compute the Ly$\alpha$ escape fraction for a subset with H$\beta$ and H$\alpha$ measurement in our sample (eight galaxies; \citealt{Tang2019}). The Ly$\alpha$ escape fraction ($f_{{\rm esc,Ly}\alpha}$) is calculated from the ratio of observed Ly$\alpha$ luminosity to the intrinsic Ly$\alpha$ luminosity ($L_{{\rm Ly}\alpha,{\rm int}}$). We derive the intrinsic Ly$\alpha$ flux from the dust-corrected H$\alpha$ luminosity assuming case B recombination ($L_{{\rm Ly}\alpha,{\rm int}}=8.7\times L_{{\rm H}\alpha,{\rm corrected}}$; e.g., \citealt{Hayes2010,Erb2014,Henry2015,Trainor2015,Jaskot2019}). The derived $f_{{\rm esc,Ly}\alpha}$ ranges from $0.05$ to $0.41$. 

The Ly$\alpha$ emission line EWs are computed from the measured Ly$\alpha$ line fluxes and the underlying continuum flux densities. Due to the lack of high S/N ($>5$) continuum measurement in our Binospec spectra, we estimate the continuum flux density using the broadband photometry in \citet{Skelton2014} catalogs. We fit the broadband fluxes from filters covering rest-frame $1250-2600$~\AA\ with a power law $f_{\lambda}\propto\lambda^{\beta}$ \citep{Calzetti1994}. Then using the fitted $f_{\lambda}-\lambda$ relation we derive the average flux density at rest-frame $1225-1250$~\AA\ as the continuum flux density \citep{Kornei2010,Stark2010}. Dividing the measured Ly$\alpha$ flux by the continuum flux density, the Ly$\alpha$ EWs of the $42$ galaxies with Ly$\alpha$ detections in our sample are from $1$~\AA\ to $136$~\AA\ with a median of $23$~\AA. 

The relationship between Ly$\alpha$ EW and [O~{\small III}]+H$\beta$ EW is shown in the right panel of Figure~\ref{fig:ssfr_mass_lyaew_o3hbew}. We find that the Ly$\alpha$ EW increases with [O~{\small III}]+H$\beta$ EW as has been shown previously (\citealt{Du2020,Tang2021b}). The median Ly$\alpha$ EW ranges from $12$~\AA\ at EW$_{{\rm [OIII]+H}\beta}=300-500$~\AA\ to $18$~\AA\ at EW$_{{\rm [OIII]+H}\beta}=500-1000$~\AA\ to $25$~\AA\ at EW$_{{\rm [OIII]+H}\beta}=1000-1500$~\AA. For galaxies with the largest [O~{\small III}]+H$\beta$ EWs in our sample ($=1500-3000$~\AA), the median Ly$\alpha$ EW is much larger ($=41$~\AA). Galaxies with such extremely large [O~{\small III}]+H$\beta$ EW become much more common at $z>6$ \citep[e.g.,][]{Smit2015,Endsley2021a,Endsley2023,Boyett2024}. While there are also moderately strong Ly$\alpha$ emitters (EW$_{{\rm Ly}\alpha}=3-40$~\AA) at EW$_{{\rm [OIII]+H}\beta}>1500$~\AA, we starting seeing extremely strong Ly$\alpha$ (EW$_{{\rm Ly}\alpha}=70-150$~\AA) among these very intense optical line emitters. Overall, we have obtained high-resolution Ly$\alpha$ spectra for sources spanning [O~{\small III}]+H$\beta$ EW $=300-3000$~\AA, allowing investigation of the Ly$\alpha$ profiles in galaxies with similar properties as those found in the reionization-era. 

\subsection{Ly$\alpha$ Profile Measurements} \label{sec:measurement}


\begin{figure*}
\includegraphics[width=\linewidth]{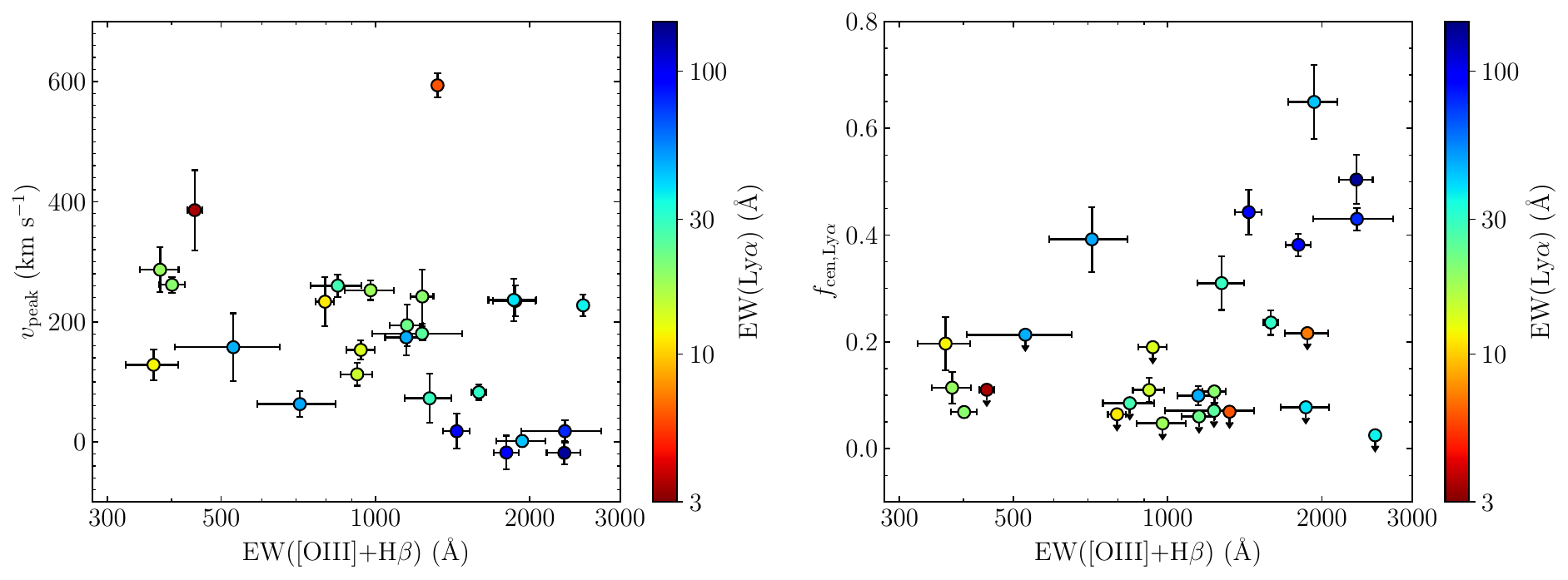}
\caption{Ly$\alpha$ peak velocity offset ($v_{\rm peak}$; left panel) and Ly$\alpha$ central escape fraction ($f_{{\rm cen,Ly}\alpha}$; right panel) as a function of [O~{\scriptsize III}]+H$\beta$ EW for the $26$ EELGs at $z=2.1-3.4$ with systemic redshift measurement in our sample. Data are color-coded by Ly$\alpha$ EWs. Galaxies with small $v_{\rm peak}$ ($<100$~km~s$^{-1}$) and large $f_{{\rm cen,Ly}\alpha}$ ($>0.2$) are found to have the largest [O~{\scriptsize III}]+H$\beta$ EWs ($>1200$~\AA) and large Ly$\alpha$ EWs ($>40$~\AA).}
\label{fig:vpeak_fcen_o3hbew}
\end{figure*}

In this section, we characterize the Ly$\alpha$ line profiles of our sample of EELGs. Of the $42$ sources with Ly$\alpha$ in emission, thirty show a faint blue peak and brigher red peak. One object presents a more complex Ly$\alpha$ profile (UDS-07665) with three peaks (Figure~\ref{fig:lya_spec1}), similar to that seen in the Sunburst arc \citep{Rivera-Thorsen2017}. The remaining $11$ EELGs in our sample show single-peaked Ly$\alpha$ lines. 

We first consider the velocity offset of the Ly$\alpha$ peak redshift and the systemic redshift ($v_{\rm peak}$). The Ly$\alpha$ velocity offset is sensitive to the column density of H~{\small I} on the far side of the galaxy \citep[e.g.,][]{Verhamme2006,Verhamme2015}, providing an indirect probe of the transmission of Ly$\alpha$ photons, assuming a symmetric distribution of H~{\small I} \citep[e.g.,][]{Verhamme2008}. Computing $v_{\rm peak}$ requires precise measurement of the systemic redshift ($z_{\rm sys}$). At $z=2.1-3.4$, the red end of Binospec grating used in this work cuts off at $\lesssim$ rest-frame $1650$~\AA. With this wavelength coverage, it is not possible to recover systemic redshifts (via non-resonant UV emission lines; i.e., O~{\small III}], C~{\small III}]) using the Binospec spectra. Instead, we use the bright (and high S/N) [O~{\small III}] or H$\alpha$ emission lines detected in near-infrared (NIR) spectra with relatively high resolution ($R>1000$) to derive systemic redshifts. Out of the $42$ EELGs at $z=2.1-3.4$, [O~{\small III}] or H$\alpha$ emission have been detected in $26$ galaxies from our ground-based spectroscopic campaign \citep{Tang2019}. 

To ensure our systemic redshift measurements are robust, we examine the consistency of systemic redshifts derived between Binospec and NIR spectra. Although we are not able to derive systemic redshifts for our $z=2.1-3.4$ objects from our Binospec spectra, we take advantage of lower redshift ($z=1.4-1.7$) sources with spectra obtained from the same Binospec and NIR multi-slit observations. For these $z=1.4-1.7$ galaxies, we derive systemic redshifts based on resolved C~{\small III}] detections in Binospec spectra, and [O~{\small III}] or H$\alpha$ detections in NIR spectra \citep{Tang2019} separately. We find that the systemic redshift derived from Binospec and NIR spectra are consistent, with a systematic uncertainty of $\simeq40$~km~s$^{-1}$, similar to the Binospec spectra resolution. This suggests the systemic redshifts derived from NIR spectra in \citet{Tang2019} should be sufficiently robust in inferring the Ly$\alpha$ peak velocity offsets for our $z=2.1-3.4$ EELGs. 

We derive the Ly$\alpha$ peak offset by measuring the shift of the Ly$\alpha$ profile maximum with respect to the systemic redshift. For the $26$ EELGs at $z=2.1-3.4$ with systemic redshift measurements, their Ly$\alpha$ peak offsets range from $v_{\rm peak}=-18$~km~s$^{-1}$ to $594$~km~s$^{-1}$. In the left panel of Figure~\ref{fig:vpeak_fcen_o3hbew}, we plot Ly$\alpha$ peak offset as a function of [O~{\small III}]+H$\beta$ EW.

We also measure the velocity separation between the blue and red Ly$\alpha$ peaks ($S_{\rm peak}$), which has been widely studied in literature and has been shown to correlate with $N_{\rm HI}$ and the ionizing photon escape fraction \citep[e.g.,][]{Verhamme2015,Izotov2018,Izotov2021,Flury2022,Hu2023,Pahl2024}. For the $30$ double-peaked Ly$\alpha$ emitters in our sample, the Ly$\alpha$ peak separations range from $316$~km~s$^{-1}$ to $846$~km~s$^{-1}$. For the triple-peaked Ly$\alpha$ emitter UDS-07665, we measure the separation between the two peaks blueward and redward the central peak, resulting in $S_{\rm peak}=582$~km~s$^{-1}$. However, eleven out of the forty-two sources in our sample only show a single-peaked Ly$\alpha$ profile, preventing us from accurately measuring their peak separations. These objects may have faint and hence undetected blue peaks, or very small peak separations which are not resolved in the spectra. To avoid introducing any bias into our results, we will primarily focus on the Ly$\alpha$ peak offset $v_{\rm peak}$ in this work, although we will also briefly discuss our peak separation $S_{\rm peak}$ measurements. 


\begin{longrotatetable}
\begin{deluxetable*}{ccccccccccccc}
\tablecaption{Ly$\alpha$ properties of the $42$ galaxies with Ly$\alpha$ emission detection in our sample. Systemic redshifts ($z_{\rm sys}$) are computed by fitting [O~{\scriptsize III}]~$\lambda5007$ or H$\alpha$ lines. For the $30$ objects with double-peaked Ly$\alpha$ lines, the peak wavelengths and fluxes of blue and red peak Ly$\alpha$ are shown in $\lambda_{\rm blue}$, $\lambda_{\rm red}$, F$_{{\rm Ly}\alpha,{\rm blue}}$, and F$_{{\rm Ly}\alpha,{\rm red}}$. For the $11$ objects with single-peaked Ly$\alpha$ emission, the central wavelengths and total fluxes are shown in $\lambda_{\rm red}$ and F$_{{\rm Ly}\alpha,{\rm red}}$. For UDS-07665 which shows triple-peaked Ly$\alpha$, we show its central peak wavelength in $\lambda_{\rm red}$ and the total flux of the central and red peaks in F$_{{\rm Ly}\alpha,{\rm red}}$. Ly$\alpha$ EWs are given as the total Ly$\alpha$ EWs. FWHMs are computed for the entire Ly$\alpha$ emission for single-peaked Ly$\alpha$ emitting galaxies, and for the red-peak Ly$\alpha$ emission for double-peaked Ly$\alpha$ emitting galaxies. The flux units are erg~s$^{-1}$~cm$^{-2}$ (cgs).}
\tabletypesize{\scriptsize}
\tablehead{
Target ID & R.A. & Decl. & $z_{\rm sys}$ & $\lambda_{\rm blue}$ & $\lambda_{\rm red}$ & F$_{{\rm Ly}\alpha,{\rm blue}}$ & F$_{{\rm Ly}\alpha,{\rm red}}$ & EW$_{{\rm Ly}\alpha}$ & FWHM & $v_{\rm peak}$ & $f_{{\rm cen,Ly}\alpha}$ & EW$_{{\rm [OIII]+H}\beta}$ \\
 & (hh:mm:ss) & (dd:mm:ss) & & (\AA) & (\AA) & ($\times10^{-18}$~cgs) & ($\times10^{-18}$~cgs) & (\AA) & (km~s$^{-1}$) & (km~s$^{-1}$) & & (\AA) 
}
\startdata
UDS-19835 & 02:17:01.94 & -05:12:35.47 & ... & $3880.75$ & $3888.33$ & $ 13.28\pm 2.59$ & $ 26.18\pm 3.36$ & $  8.8\pm 0.9$ & $333\pm32$ & ... & ... & $ 387\pm 29$ \\
UDS-32530 & 02:17:04.81 & -05:10:02.64 & $2.1512$ & $3828.39$ & $3832.46$ & $  2.10\pm 4.11$ & $ 31.74\pm 6.08$ & $ 11.7\pm 2.5$ & $385\pm56$ & $128\pm26$ & $0.20\pm0.05$ & $ 369\pm 43$ \\
UDS-09067 & 02:17:01.48 & -05:14:45.36 & $3.2286$ & $5129.98$ & $5144.48$ & $ 18.86\pm 4.75$ & $ 58.38\pm 5.03$ & $ 36.1\pm 3.2$ & $417\pm27$ & $227\pm18$ & $<0.03$ & $2541\pm 63$ \\
UDS-08078 & 02:17:02.74 & -05:14:57.50 & $3.2277$ & ... & $5149.66$ & ... & $ 22.34\pm 3.22$ & $  6.0\pm 0.9$ & $414\pm45$ & $593\pm20$ & $<0.07$ & $1321\pm 30$ \\
UDS-22025 & 02:17:06.36 & -05:12:06.88 & ... & ... & $3849.48$ & ... & $ 16.27\pm 1.94$ & $ 12.9\pm 1.5$ & $542\pm95$ & ... & ... & $ 476\pm 71$ \\
UDS-27576 & 02:17:08.11 & -05:10:59.82 & ... & $3860.37$ & $3867.61$ & $ 14.57\pm 7.71$ & $ 33.27\pm 6.63$ & $ 23.4\pm 5.0$ & $349\pm52$ & ... & ... & $ 452\pm 78$ \\
UDS-21302 & 02:17:07.94 & -05:12:15.51 & $2.1804$ & $3862.43$ & $3868.35$ & $  8.46\pm 3.35$ & $ 10.50\pm 4.69$ & $ 46.6\pm14.1$ & $288\pm95$ & $157\pm56$ & $<0.21$ & $ 528\pm122$ \\
UDS-11049 & 02:17:07.14 & -05:14:19.88 & ... & $3932.50$ & $3937.91$ & $  6.67\pm 1.19$ & $ 44.58\pm13.02$ & $ 35.5\pm 9.0$ & $263\pm28$ & ... & ... & $ 672\pm 80$ \\
UDS-33286 & 02:17:12.40 & -05:09:52.91 & ... & $3995.26$ & $4003.77$ & $  1.85\pm 0.96$ & $ 14.13\pm 4.42$ & $  2.5\pm 0.7$ & $266\pm34$ & ... & ... & $ 620\pm 21$ \\
UDS-32404 & 02:17:13.54 & -05:10:02.79 & ... & ... & $3816.50$ & ... & $ 14.04\pm 7.75$ & $ 10.7\pm 5.9$ & $280\pm117$ & ... & ... & $ 528\pm 53$ \\
UDS-37633 & 02:17:16.05 & -05:08:57.44 & ... & $3754.76$ & $3759.57$ & $ 19.10\pm 9.46$ & $ 57.09\pm15.72$ & $ 32.5\pm 7.8$ & $331\pm32$ & ... & ... & $ 926\pm 65$ \\
UDS-23016 & 02:17:15.48 & -05:11:55.36 & $2.1541$ & $3830.98$ & $3838.01$ & $ 11.75\pm 5.72$ & $ 39.79\pm 8.90$ & $ 18.7\pm 3.8$ & $504\pm84$ & $286\pm37$ & $0.11\pm0.03$ & $ 380\pm 33$ \\
UDS-21724 & 02:17:20.01 & -05:12:10.62 & $3.2279$ & $5128.57$ & $5141.15$ & $  5.68\pm 1.61$ & $ 28.17\pm 2.29$ & $ 30.2\pm 2.5$ & $410\pm25$ & $ 82\pm12$ & $0.24\pm0.02$ & $1591\pm 51$ \\
UDS-14097 & 02:17:22.44 & -05:13:42.89 & $2.1565$ & $3833.94$ & $3839.49$ & $  8.52\pm 2.55$ & $ 43.36\pm 3.54$ & $ 47.3\pm 4.0$ & $258\pm16$ & $174\pm29$ & $0.10\pm0.02$ & $1148\pm104$ \\
UDS-10245 & 02:17:22.93 & -05:14:30.63 & $2.2995$ & $4006.73$ & $4014.48$ & $  8.42\pm 3.17$ & $ 39.98\pm 6.89$ & $ 18.0\pm 2.8$ & $289\pm27$ & $252\pm16$ & $<0.05$ & $ 978\pm107$ \\
UDS-10805 & 02:17:23.71 & -05:14:22.97 & $2.2924$ & $3999.70$ & $4004.88$ & $ 12.20\pm 3.30$ & $ 19.70\pm 3.03$ & $ 24.8\pm 3.5$ & $129\pm14$ & $180\pm11$ & $<0.07$ & $1232\pm245$ \\
UDS-15533 & 02:17:26.08 & -05:13:25.28 & $2.1572$ & $3832.82$ & $3840.60$ & $ 16.79\pm 5.50$ & $ 36.39\pm 6.72$ & $ 22.7\pm 3.7$ & $342\pm52$ & $194\pm35$ & $<0.06$ & $1152\pm 87$ \\
UDS-14425 & 02:17:26.38 & -05:13:40.64 & $2.2103$ & $3897.36$ & $3906.07$ & $ 21.71\pm 4.28$ & $ 68.22\pm 5.70$ & $ 19.7\pm 1.6$ & $339\pm20$ & $261\pm13$ & $0.07\pm0.01$ & $ 401\pm 23$ \\
UDS-27411 & 02:17:29.13 & -05:11:01.46 & $2.2096$ & ... & $3902.76$ & ... & $ 28.72\pm 4.03$ & $ 29.4\pm 4.1$ & $316\pm33$ & $ 72\pm40$ & $0.31\pm0.05$ & $1275\pm133$ \\
UDS-18860 & 02:17:29.26 & -05:12:45.54 & $2.2103$ & ... & $3905.72$ & ... & $  9.54\pm 2.49$ & $  7.1\pm 1.8$ & $191\pm35$ & $234\pm25$ & $<0.22$ & $1874\pm180$ \\
UDS-04725 & 02:17:27.93 & -05:15:37.62 & ... & $4022.27$ & $4027.08$ & $  8.97\pm 4.18$ & $ 27.28\pm 4.33$ & $ 22.4\pm 3.7$ & $228\pm25$ & ... & ... & $1105\pm115$ \\
UDS-27009 & 02:17:33.88 & -05:11:06.40 & ... & ... & $3900.54$ & ... & $ 72.30\pm10.87$ & $ 60.1\pm 9.0$ & $319\pm36$ & ... & ... & $1135\pm107$ \\
UDS-11394 & 02:17:32.18 & -05:14:16.25 & $2.1831$ & $3865.76$ & $3869.83$ & $ 31.09\pm 8.88$ & $ 96.78\pm 8.23$ & $ 97.0\pm 9.2$ & $190\pm15$ & $ 17\pm28$ & $0.44\pm0.04$ & $1440\pm 88$ \\
UDS-13231 & 02:17:33.19 & -05:13:53.71 & ... & ... & $4006.36$ & ... & $ 12.34\pm 5.00$ & $  5.4\pm 2.2$ & $228\pm72$ & ... & ... & $ 342\pm 36$ \\
UDS-27151 & 02:17:36.14 & -05:11:06.18 & $2.1538$ & $3828.76$ & $3835.42$ & $ 15.09\pm 8.23$ & $ 59.66\pm 8.77$ & $ 14.6\pm 2.3$ & $344\pm37$ & $112\pm18$ & $0.11\pm0.02$ & $ 921\pm 65$ \\
UDS-07665 & 02:17:33.78 & -05:15:02.85 & $2.2964$ & $4001.92$ & $4007.10$ & $ 12.06\pm 2.52$ & $ 93.85\pm 4.74$ & $ 96.6\pm 4.9$ & ... & $-17\pm28$ & $0.38\pm0.02$ & $1800\pm101$ \\
UDS-13128 & 02:17:35.23 & -05:13:54.79 & ... & $3836.90$ & $3841.34$ & $ 15.96\pm12.35$ & $ 13.31\pm 7.53$ & $ 50.9\pm25.1$ & $145\pm57$ & ... & ... & $ 430\pm100$ \\
UDS-05713 & 02:17:36.23 & -05:15:26.38 & ... & ... & $4048.17$ & ... & $  9.02\pm 1.51$ & $  7.2\pm 1.2$ & ... & ... & ... & $ 361\pm 43$ \\
UDS-22650 & 02:17:39.96 & -05:11:59.00 & $2.1708$ & $3853.18$ & $3857.99$ & $  3.74\pm 2.29$ & $ 26.80\pm 5.09$ & $ 27.8\pm 5.1$ & $256\pm31$ & $260\pm18$ & $<0.09$ & $ 844\pm 96$ \\
UDS-11222 & 02:17:39.10 & -05:14:17.88 & $2.1528$ & $3824.82$ & $3833.57$ & $  8.84\pm 3.78$ & $ 30.11\pm 3.89$ & $ 47.9\pm 6.7$ & $221\pm21$ & $ 63\pm21$ & $0.39\pm0.06$ & $ 712\pm124$ \\
UDS-29766 & 02:17:43.46 & -05:10:33.45 & $2.3024$ & $4007.06$ & $4014.87$ & $  6.39\pm 2.42$ & $ 74.06\pm 2.70$ & $ 78.6\pm 3.5$ & $290\pm 7$ & $ 18\pm18$ & $0.43\pm0.02$ & $2341\pm418$ \\
UDS-23682 & 02:17:43.01 & -05:11:47.56 & ... & ... & $3939.78$ & ... & $  2.39\pm 0.99$ & $  0.9\pm 0.4$ & ... & ... & ... & $ 488\pm 51$ \\
UDS-27040 & 02:17:44.29 & -05:11:06.22 & $2.1921$ & ... & $3880.56$ & ... & $103.60\pm11.08$ & $ 43.2\pm 4.6$ & $201\pm 5$ & $  1\pm 9$ & $0.65\pm0.07$ & $1932\pm213$ \\
UDS-19167 & 02:17:43.54 & -05:12:43.61 & $2.1847$ & $3866.13$ & $3871.31$ & $ 67.63\pm 7.50$ & $352.90\pm37.72$ & $136.1\pm12.4$ & $234\pm 5$ & $-18\pm18$ & $0.50\pm0.05$ & $2335\pm178$ \\
UDS-29927 & 02:17:46.26 & -05:10:32.68 & $2.1987$ & $3880.06$ & $3890.55$ & $  2.93\pm 3.10$ & $ 16.97\pm 6.77$ & $ 13.5\pm 5.0$ & $181\pm27$ & $153\pm16$ & $<0.19$ & $ 936\pm 60$ \\
UDS-24183 & 02:17:47.40 & -05:11:41.45 & $2.2448$ & $3937.06$ & $3947.79$ & $  8.36\pm 5.84$ & $ 25.28\pm 7.29$ & $ 19.8\pm 5.5$ & $365\pm81$ & $242\pm44$ & $<0.11$ & $1234\pm 63$ \\
UDS-20810 & 02:17:48.02 & -05:12:20.91 & $2.2559$ & $3953.08$ & $3961.22$ & $ 14.80\pm 4.37$ & $ 10.55\pm 3.76$ & $ 38.4\pm 8.7$ & $252\pm72$ & $236\pm35$ & $<0.08$ & $1861\pm201$ \\
UDS-19130 & 02:17:51.10 & -05:12:45.17 & $2.2941$ & ... & $4009.69$ & ... & $ 18.93\pm 5.95$ & $  3.4\pm 1.1$ & $460\pm123$ & $385\pm66$ & $<0.11$ & $ 444\pm 15$ \\
UDS-25351 & 02:17:54.23 & -05:11:27.56 & ... & $3854.66$ & $3859.10$ & $  3.16\pm 2.43$ & $ 29.62\pm 4.85$ & $ 23.2\pm 3.8$ & $311\pm31$ & ... & ... & $ 399\pm 47$ \\
UDS-06274 & 02:17:52.31 & -05:15:20.26 & ... & $4971.32$ & $4983.90$ & $  1.62\pm 1.10$ & $ 13.07\pm 3.38$ & $  8.1\pm 2.0$ & $433\pm78$ & ... & ... & $ 550\pm 75$ \\
UDS-08728 & 02:17:56.20 & -05:14:49.26 & ... & $3942.35$ & $3950.86$ & $  6.50\pm 3.72$ & $ 47.72\pm 4.11$ & $ 43.1\pm 4.4$ & $254\pm24$ & ... & ... & $ 686\pm 69$ \\
UDS-08964 & 02:17:57.38 & -05:14:45.59 & $2.2462$ & $3941.24$ & $3949.38$ & $ 13.72\pm 4.57$ & $ 20.04\pm 4.94$ & $ 11.2\pm 2.2$ & $335\pm39$ & $233\pm40$ & $<0.06$ & $ 797\pm 32$ \\
\enddata
\label{tab:lya_profile}
\end{deluxetable*}
\end{longrotatetable}

We next consider the fraction of Ly$\alpha$ emission within $\pm100$~km~s$^{-1}$ of the systemic velocity (the Ly$\alpha$ `central escape fraction', $f_{{\rm cen,Ly}\alpha}$, as defined in \citealt{Naidu2022}) in order to constrain the H~{\small I} covering fraction. In a clumpy H~{\small I} distribution, Ly$\alpha$ photons can escape directly through low-opacity ($\tau\ll1$, or equivalently $N_{\rm HI}\lesssim10^{13}$~cm$^{-2}$; e.g., \citealt{Dijkstra2016a,Ouchi2020}) channels, as may be expected if a subset of the massive stars are (partially) covered by low density H~{\small I} with highly ionized sightlines \citep[e.g.,][]{Gazagnes2020,Ma2020}. This results in a significant fraction of the Ly$\alpha$ line emerging at the systemic redshift, as shown in both simulations \citep[e.g.,][]{Behrens2014,Verhamme2015,Dijkstra2016b} and observations \citep[e.g.,][]{Rivera-Thorsen2017}. When computing $f_{{\rm cen,Ly}\alpha}$ ($=$ Ly$\alpha$ flux at $\pm100$~km~s$^{-1}/$ total Ly$\alpha$ flux), we choose the same central velocity window ($\pm100$~km~s$^{-1}$) as in \citet{Naidu2022} since both samples have similar spectral resolution ($R\simeq4000$) around Ly$\alpha$ \citep{Matthee2021}. We show $f_{{\rm cen,Ly}\alpha}$ as a function 
of [O~{\small III}]+H$\beta$ EW in the right panel of Figure~\ref{fig:vpeak_fcen_o3hbew}. 
For the $26$ EELGs with $z_{\rm sys}$ in our sample, the measured Ly$\alpha$ central escape fraction ranges from $f_{{\rm cen,Ly}\alpha}<0.02$ ($3\sigma$ upper limit) to $f_{{\rm cen,Ly}\alpha}=0.65$.

The flux ratio of blue to red Ly$\alpha$ peaks (so called ``blue-to-red ratio'') also provides constrains the H~{\small I} and dust content. Since the blueshifted Ly$\alpha$ emission faces significant scattering through the near side of the galaxy, a larger blue-to-red flux ratio may imply a low $N_{\rm HI}$ and less dust. We measure the blue-to-red Ly$\alpha$ ratio for the $30$ EELGs with double-peaked Ly$\alpha$ emission in our sample. The median blue-to-red ratio of these $30$ objects is $0.29$, which is consistent with the average blue-to-red flux ratio of Ly$\alpha$ emission lines of Ly$\alpha$ emitting galaxies at $z\sim2-3$ \citep[e.g.,][]{Trainor2015,Hayes2021,Matthee2021}. 

Finally, we quantify the full width at half maximum (FWHM) of the Ly$\alpha$ profile for our EELGs to provide a measurement of the width of red Ly$\alpha$ damping wings. The FWHM is computed by subtracting the instrument resolution in quadrature from the observed FWHM. For the $30$ double-peaked Ly$\alpha$ emitters in our sample, the median FWHMs of blue-peak and red-peak Ly$\alpha$ emission are $264$~km~s$^{-1}$ and $290$~km~s$^{-1}$, respectively. For the $11$ single-peaked Ly$\alpha$ emitters, the median FWHM is $316$~km~s$^{-1}$. Since the blueshifted Ly$\alpha$ emission is more likely to be affected by the residual neutral hydrogen in the IGM, we quote FWHM as the FWHM of the red-peak emission for double-peaked Ly$\alpha$ emitters, or the FWHM of the entire emission for single-peaked Ly$\alpha$ emitters in the following. We summarize the Ly$\alpha$ profile measurements of our $z=2.1-3.4$ EELGs in Table~\ref{tab:lya_profile}.


\section{L\lowercase{y}$\alpha$ in Extreme [O~{\small III}] Emitters} \label{sec:result}

The spectra described in the previous section allow us to characterize the Ly$\alpha$ profiles in $z\simeq2-3$ galaxies with extremely large [O~{\small III}]+H$\beta$ EWs, a population of low mass galaxies with large sSFR, similar to what is commonly seen in the reionization era. While Ly$\alpha$ profiles are potentially useful for insight into Lyman continuum (LyC) escape in this population \citep[e.g.,][]{Verhamme2015,Verhamme2017,Izotov2018,Rivera-Thorsen2019,Gazagnes2020,Naidu2022,Flury2022,Pahl2024}, our primary focus in this paper is on developing a baseline for interpreting the database of $z\gtrsim7$ line profiles now emerging from {\it JWST} observations. We begin in Section~\ref{sec:high_fcen} by describing the properties of galaxies with significant Ly$\alpha$ transmission at line center, potentially signalling very low density (and highly ionized) channels facilitating direct escape of Ly$\alpha$. We then describe the range of Ly$\alpha$ profiles seen in galaxies as a function of [O~{\small III}]+H$\beta$ EW, first considering values that are typical of the reionization era ($400-1200$~\AA; Section~\ref{sec:moderate_eelg}). We close by discussing the Ly$\alpha$ profiles of $z\simeq2-3$ galaxies with [O~{\small III}]+H$\beta$ EW $>1200$~\AA\ (Section~\ref{sec:strong_eelg}).

\subsection{Ly$\alpha$ Profiles with Large Central Escape Fractions} \label{sec:high_fcen}


\begin{figure*}
\includegraphics[width=\linewidth]{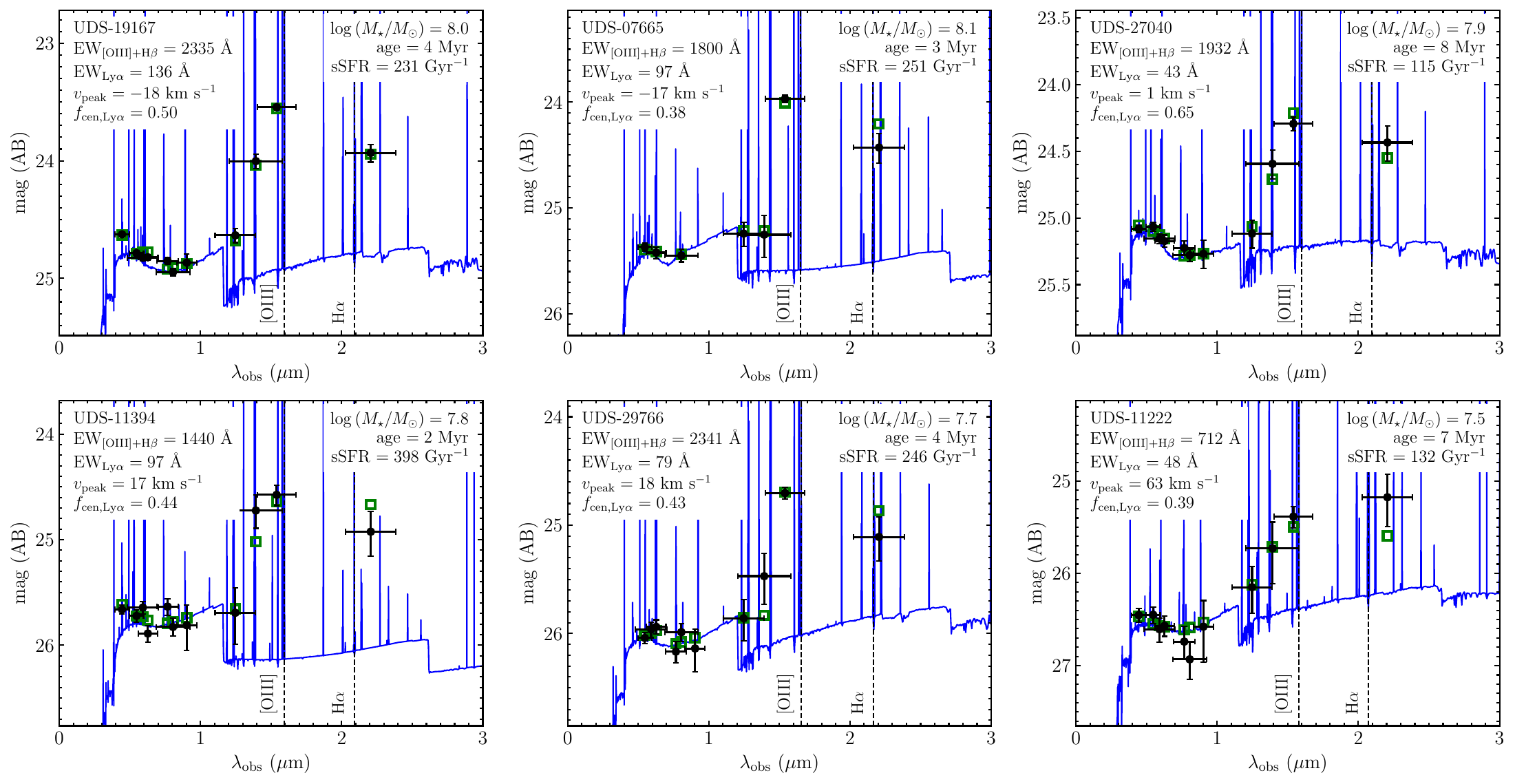}
\caption{Spectral energy distributions (SEDs) of the six galaxies with high Ly$\alpha$ central escape fraction ($f_{{\rm cen,Ly}\alpha}\gtrsim0.4$) in our $z=2.1-3.4$ EELG sample. Observed broadband photometry is shown as solid black circles. The best-fit SED models inferred from {\tt BEAGLE} are plotted by solid blue lines, and synthetic photometry is presented by open green squares. These objects are characterized by extremely young stellar populations ($2-8$~Myr assuming CSFH) and low stellar masses ($3\times10^7-1\times10^8\ M_{\odot}$).}
\label{fig:lya_sed}
\end{figure*}

Recently attention has focused on the subset of galaxies with Ly$\alpha$ photons directly escaping near ($\pm100$~km~s$^{-1}$) line center \citep[e.g.,][]{Rivera-Thorsen2017,Naidu2022}. One possible explanation for such line profiles is partial coverage of neutral hydrogen, with holes that allow transmission of Ly$\alpha$ at the systemic redshift. \citet{Naidu2022} have characterized the central flux fraction of Ly$\alpha$ ($f_{{\rm cen,Ly}\alpha}$; see Section~\ref{sec:measurement} for definition) for a sample of $z=0-4$ galaxies with LyC leakage, with the results showing large $f_{{\rm cen,Ly}\alpha}$ values ($0.1-0.4$) in the strongest leakers \citep[e.g.,][]{Naidu2017,Rivera-Thorsen2019,Izotov2021,Matthee2021}. Analysis of $z\simeq2$ galaxies selected on strong Ly$\alpha$ also reveals a subset with emission near line center, with values reaching as high as $f_{{\rm cen,Ly}\alpha}=0.2-0.5$ \citep{Naidu2022}. Interest in this subclass of the Ly$\alpha$ emitter population is partially driven by their potential as LyC leaking candidates given the very low H~{\small I} column density implied by direct Ly$\alpha$ escape\footnote{Large values of $f_{{\rm cen,Ly}\alpha}$ may alternatively be driven by very small separation of the blue and red peaks (i.e., $S_{\rm peak}\lesssim100$~km~s$^{-1}$), with the majority of flux emerging close to line center. This case also requires low H~{\scriptsize I} column densities (albeit not as low as required for direct escape) and may be linked to significant LyC escape. We will comment on this later in the section.} \citep{Naidu2022,Choustikov2024}. Recent work has also highlighted the utility of this population for probing the IGM at $z\gtrsim7$ \citep{Saxena2023,Tang2024}. Since significant transmission near the systemic redshift is only expected in large ionized regions \citep[e.g.,][]{Mason2020}, the identification of $z\gtrsim7$ galaxies with both large $f_{{\rm cen,Ly}\alpha}$ (or small $v_{\rm peak}$) and large Ly$\alpha$ escape fractions enables constraints on the proximity of the galaxy to neutral hydrogen in the IGM \citep[e.g.,][]{Prieto-Lyon2023,Lu2024}.

The low density (and highly ionized) sightlines that facilitate Ly$\alpha$ with large $f_{{\rm cen,Ly}\alpha}$ \citep[e.g.,][]{Behrens2014,Erb2014,Verhamme2015,Dijkstra2016b} have been suggested to arise shortly after the intense bursts of star formation \citep[e.g.,][]{Smith2019,Ma2020} that appear fairly ubiquitously at $z\gtrsim6$ \citep[e.g.,][]{Labbe2013,Smit2014,Endsley2021a,Endsley2023,Topping2022,Whitler2023b}. Our sample (selected on [O~{\small III}]+H$\beta$ EW) allows us to quantify how commonly Ly$\alpha$ has large central flux fractions in low mass galaxies experiencing rapid upturns of star formation at $z\simeq2-3$. Figure~\ref{fig:lya_spec1} shows the six galaxies with the largest Ly$\alpha$ central flux fractions in our high resolution MMT spectra. The values spanned in these galaxies ($f_{{\rm cen,Ly}\alpha}=0.38-0.65$) are as large as any of the LyC leakers and Ly$\alpha$ emitters considered in \citet{Naidu2022}, potentially indicating that EELGs (at least occasionally) produce very low H~{\small I} density channels that allow direct escape of Ly$\alpha$.

The SEDs of the six large $f_{{\rm cen,Ly}\alpha}$ systems are shown in Figure~\ref{fig:lya_sed}. The UV continuum slopes of these galaxies are relatively blue (median $\beta=-2.3$), indicating low dust attenuation. The galaxies have among the largest [O~{\small III}]+H$\beta$ EWs in the sample (EW $=712-2341$~\AA, with a median EW $\simeq1900$~\AA). The light-weighted ages of the six galaxies are correspondingly young, ranging from $2$ to $8$~Myr, with a median of $4$~Myr (all assuming CSFH). The stellar masses are low, ranging from $3\times10^7$ to $1\times10^8\ M_{\odot}$. The derived sSFRs ($115$ to $400$~Gyr$^{-1}$) point to galaxies caught in the midst of a significant burst of star formation. The existence of strong and centrally-peaked Ly$\alpha$ emission in this subset of galaxies may indicate that low H~{\small I} density channels can form quickly in low mass galaxies during intense star formation episodes. This is consistent with the short timescale of forming low density channels ($\simeq1-3$~Myr) indicated from hydrodynamical simulations \citep[e.g.,][]{Ma2020,Kakiichi2021}. We will show in Section~\ref{sec:moderate_eelg} and Section~\ref{sec:strong_eelg} that these conditions only occur in a subset of the extreme [O~{\small III}] emitting population, suggesting that not all low mass galaxies have extremely low density neutral gas along our sightline during large sSFR phases. 

We expect the six galaxies in our sample with centrally prominent Ly$\alpha$ emission to have large Ly$\alpha$ EWs, both because of the effective transmission implied by the line profile and the efficient ionizing photon production implied by the young stellar population ages. This is indeed the case, with the majority showing intense Ly$\alpha$ emission (median EW $=90$~\AA). The Ly$\alpha$ escape fractions derived from H$\alpha$ (see Section~\ref{sec:spectra}) indicate larger-than-average transmission relative to typical continuum-selected galaxies \citep[e.g.,][]{Hayes2010,Erb2014,Matthee2016,Sobral2017}, with a median value of $f_{{\rm esc,Ly}\alpha}=0.22$ (ranging from $f_{{\rm esc,Ly}\alpha}=0.16$ to $0.41$). The Ly$\alpha$ escape fractions are slightly lower than those of Green Peas with similar Ly$\alpha$ EWs (median $f_{{\rm esc,Ly}\alpha}=0.3$; \citealt{Yang2017}), likely because the emission from diffuse Ly$\alpha$ halos is not fully recovered by the $1$~arcsec width Binospec slit \citep[e.g.,][]{LujanNiemeyer2022}. Yet in spite of these enhanced values, the escape fractions imply that the majority of the Ly$\alpha$ emission is not detected, as would be expected if the low H~{\small I} density gas (that permits transmission near line center) is surrounded by denser gas which does scatter Ly$\alpha$ photons. 

The Ly$\alpha$ profiles offer further insight into the mode of Ly$\alpha$ escape in galaxies with centrally-peaked Ly$\alpha$ emission. In addition to the prominent central component, we generally also see a blue peak and red tail of emission extending to higher velocities ($100-400$~km~s$^{-1}$). This confirms the suggestion that a significant fraction of the Ly$\alpha$ photons arrive via resonant scattering or backscattering through dense H~{\small I}, also indicating that the lower H~{\small I} density gas (that permit emission at systemic) is likely surrounded by denser gas (which scatters Ly$\alpha$ photons to larger velocities). 

The Ly$\alpha$ spectrum of UDS-07665 ([O~{\small III}]+H$\beta=1800$~\AA, $f_{{\rm cen,Ly}\alpha}=0.38$) provides an illustrative example. The Ly$\alpha$ profile shows three components with a central peak close to the systemic, similar to that seen in some known LyC leakers (and strong Ly$\alpha$ emitters) at $z=0-4$, including the Sunburst Arc \citep{Rivera-Thorsen2017}, J1243+4646 \citep{Izotov2018}, and {\it Ion3} \citep{Vanzella2018}. The central component in UDS-07665 has a peak velocity of $v_{\rm peak}=-17$~km~s$^{-1}$ and narrow line width (FWHM $=74$~km~s$^{-1}$). The central Ly$\alpha$ component line width is comparable to that of the H$\alpha$ emission line of UDS-07665 (FWHM $=88$~km~s$^{-1}$ after subtracting the intrument resolution in quadrature) and also those of non-resonant rest-frame optical emission lines in similarly-selected galaxies \citep[e.g.,][]{Maseda2014,Tang2022}, as would be expected if it arises via direct escape without scattering. The blue peak ($v=-405$~km~s$^{-1}$) and red peak ($v=+177$~km~s$^{-1}$) velocities are consistent with expectations for scattering through (and backscattering off of) dense outflowing neutral gas. Notably this implies that the separation between the red and blue peaks is large ($S_{\rm peak}=582$~km~s$^{-1}$), in spite of the presence of gas conditions that permit significant Ly$\alpha$ transmission at line center. This underlies why peak separation is not a sufficient criterion for identifying galaxies with very low H~{\small I} column density channels, as pointed out in \citet{Naidu2022} and \citet{Almada-Monter2024}. 

In one case (UDS-27040), we do not see clearly-defined and robustly-detected blue peak. It is possible that the blue and red peaks are unresolved in this system, leading to the appearance of a single central peak. This may be expected if young stars are uniformly covered by slowly-outflowing neutral hydrogen that has low column density yet is optically thick to Ly$\alpha$ \citep[e.g.,][]{Verhamme2015}. It is also possible that the S/N of this system is too low to recover the blue peak. In either case, this spectrum still imply at least partial coverage by low column densities of neutral hydrogen. 
For the six galaxies with large Ly$\alpha$ central escape fractions, we will use the radiative transfer models developed in \citet{Li2021,Li2022} to explore the physical conditions required to explain these line profiles in a future paper. 

\subsection{Ly$\alpha$ Profiles in Galaxies with [O~{\small III}]+H$\beta$ EW $=400-1200$~\AA} \label{sec:moderate_eelg}


\begin{figure*}
\includegraphics[width=\linewidth]{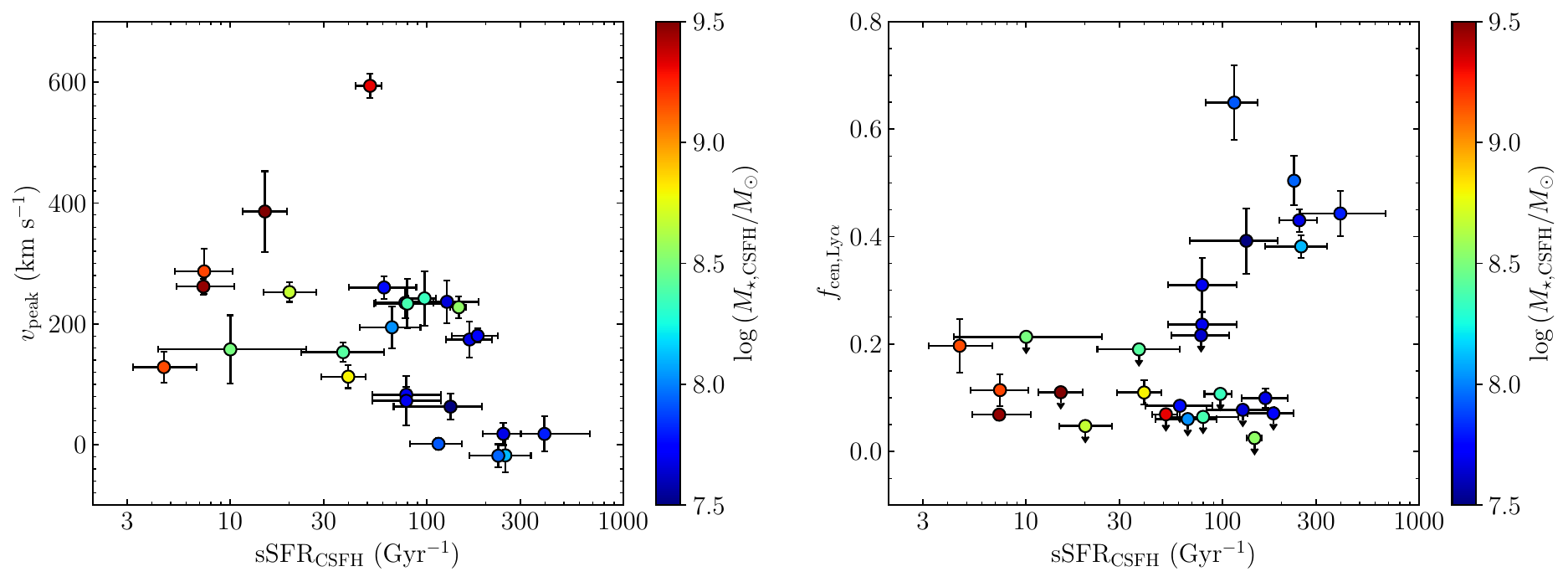}
\caption{Ly$\alpha$ peak velocity offset (left panel) and Ly$\alpha$ central escape fraction (right panel) as a function of sSFR for our EELGs at $z=2.1-3.4$. Data are color-coded by stellar mass. Galaxies with very low Ly$\alpha$ peak offsets and large Ly$\alpha$ central escape fractions are low mass systems ($M_{\star}\lesssim10^8\ M_{\odot}$) with very large sSFRs ($>100$~Gyr$^{-1}$ assuming CSFH).}
\label{fig:vpeak_fcen_ssfr}
\end{figure*}

One of the primary goals of our observations is to obtain a census of Ly$\alpha$ profiles in $z\simeq2-3$ galaxies with large [O~{\small III}]+H$\beta$ EWs. We begin by considering the $23$ objects in our high resolution sample with [O~{\small III}]+H$\beta$ EW $=400-1200$~\AA\ (with $11$ of these having systemic redshift measurements). This range is fairly typical of galaxies in the reionization era, where the median ($25-75$th percentile) [O~{\small III}]+H$\beta$ EW is $780$~\AA\ ($500-1220$~\AA; \citealt{Endsley2023}). The UV absolute magnitudes range from M$_{\rm UV}=-21$ to $-18$, similar to typical galaxies at $z\gtrsim6$ \citep[e.g.,][]{Finkelstein2015,Ishigaki2018,Bouwens2021}. The UV continuum slopes are fairly blue (median $\beta=-2.2$), signalling modest attenuation from dust. The stellar masses are found to be relatively low (median $=2.7\times10^8\ M_\odot$), and the light-weighted ages are found to be correspondingly young (median $26$~Myr). The large [O~{\small III}]+H$\beta$ EWs in this subset of our sample are primarily driven by large sSFR (median $=40$~Gyr$^{-1}$), signalling a recent upturn in star formation. These properties are all similar to what is seen in most $z\gtrsim6$ galaxies.

Our spectra reveal a diverse set of Ly$\alpha$ profiles in $z\simeq2-3$ galaxies with [O~{\small III}]+H$\beta$ EW $=400-1200$~\AA. The Ly$\alpha$ EWs range from weak ($1$~\AA) to strong ($60$~\AA) with a median value of $22$~\AA\ (see also \citealt{Du2020,Tang2021b}). The Ly$\alpha$ velocity offsets also span a wide range, from $v_{\rm peak}=63$~km~s$^{-1}$ to $386$~km~s$^{-1}$ with a median value of $194$~km~s$^{-1}$ (see examples in Figure~\ref{fig:lya_spec2}). In spite of the large sSFR and reasonably low masses, most of these galaxies show very little Ly$\alpha$ emission at line center (median with $f_{{\rm cen,Ly}\alpha}=0.1$), suggesting direct escape of Ly$\alpha$ is not ubiquitous in $z\simeq2-3$ galaxies with [O~{\small III}]+H$\beta$ EW $=400-1200$~\AA. Instead, Ly$\alpha$ photons mostly escape through backscattering and resonant scattering through the outflowing H~{\small I}, as is common in most Ly$\alpha$ emitters with lower [O~{\small III}]+H$\beta$ EWs \citep[e.g.,][]{Shapley2003,Jones2013,Ouchi2020}. The redshifted component of the Ly$\alpha$ lines are reasonably broad, with FWHM ranging between $146$ and $542$~km~s$^{-1}$ with median $=289$~km~s$^{-1}$). We occasionally detect Ly$\alpha$ flux redshifted to very high velocities. In $5$ of $11$ galaxies with [O~{\small III}]+H$\beta$ EW $=400-1200$~\AA\ and systemic redshift measurements, we detect more than $10\%$ of the line flux redshifted to $500-1000$~km~s$^{-1}$. This high velocity emission is important for visibility in the reionization era, as we will discuss in Section~\ref{sec:discussion}.

We detect blue peaks in $18$ of the $23$ $z\simeq2-3$ galaxies with [O~{\small III}]+H$\beta$ EW $=400-1200$~\AA, reflecting line photons that have resonantly scattered through the near side of the outflowing H~{\small I} (see \citealt{Ouchi2020} for a review). On average, we find $21\%$ of the total Ly$\alpha$ flux is in the blue peak, with typical velocities between $-110$ and $-650$~km~s$^{-1}$ relative to line center. We also find $12\%$ of the Ly$\alpha$ flux blueshifted to very high velocities ($-1000$ to $-500$~km~s$^{-1}$). The velocity shift of the blue peak is generally larger than that of the red peak, as expected for transfer through outflowing gas \citep[e.g.,][]{Verhamme2006,Verhamme2015,Orlitova2018,Ouchi2020}. In one case, (UDS-08964), we see a blue peak ($v_{\rm blue}=-385$~km~s$^{-1}$) that is roughly as strong as the red peak (see Figure~\ref{fig:lya_spec2}), with a blue-to-red peak flux ratio of $0.68\pm0.28$. This may reflect transfer through slow-moving H~{\small I} \citep[e.g.,][]{Verhamme2006,Verhamme2015,Li2022}, as might be expected if outflows are weak in this galaxy. Regardless of the origin, the blue peak flux fractions are important for interpreting the evolving Ly$\alpha$ EW distribution at $z\gtrsim6$. As we approach the reionization era, we expect the blue peaks to be strongly attenuated by the IGM \citep[e.g.,][]{Hayes2021}, leading to reduction of Ly$\alpha$ EWs relative to our measurements at $z\simeq2-3$. We will quantify this in Section~\ref{sec:discussion}. The peak separations of the $18$ galaxies with [O~{\small III}]+H$\beta$ EW $=400-1200$~\AA\ with double-peak Ly$\alpha$ are large, with a median of $S_{\rm peak}=571$~km~s$^{-1}$. Together with the negligible $f_{{\rm cen,Ly}\alpha}$, these indicate that H~{\small II} regions in most of the [O~{\small III}]+H$\beta$ EW $=400-1200$~\AA\ galaxies are at least partially covered by dense H~{\small I} \citep[e.g.,][]{Verhamme2015,Li2022}.

While most galaxies with [O~{\small III}]+H$\beta$ EW $=400-1200$~\AA\ appear to not have H~{\small I} conditions that facilitate significant Ly$\alpha$ escape at line center, we do identify one galaxy in this sample with large Ly$\alpha$ central flux fractions. UDS-11222 ([O~{\small III}]+H$\beta$ EW $=712$~\AA, Ly$\alpha$ EW $=48$~\AA) has a Ly$\alpha$ profile with $f_{{\rm cen,Ly}\alpha}=0.39$, more than twice what is typical in this [O~{\small III}]+H$\beta$ EW range. UDS-11222 stands out as having the lowest stellar mass ($3\times10^7\ M_\odot$ for CSFH) and the faintest M$_{\rm UV}$ ($-18.3$) in the entire $z\simeq2-3$ sample presented in this paper. The relatively low [O~{\small III}]+H$\beta$ EW of UDS-11222 (comparing to other systems with large Ly$\alpha$ central flux fractions) may be due to a lower metallicity or a more recently-declining star formation history at fainter M$_{\rm UV}$ \citep[e.g.,][]{Endsley2023b}. SED fitting results also show that UDS-11222 has the largest sSFR ($=139$~Gyr$^{-1}$) among those with [O~{\small III}]+H$\beta$ EW $=400-1200$~\AA, suggesting it is experiencing a burst of star formation that is more typical of galaxies with [O~{\small III}]+H$\beta$ EW $>1200$~\AA. The presence of bursts in low mass galaxies may help disrupt the neutral gas in galaxies, creating highly ionized channels that allow Ly$\alpha$ to escape directly \citep[e.g.,][]{Kimm2019,Ma2020}. In the next subsection, we explore whether such low density sightlines are more common in the galaxies in our sample with [O~{\small III}]+H$\beta$ EW $>1200$~\AA.

\subsection{Ly$\alpha$ Profiles in Galaxies with [O~{\small III}]+H$\beta$ EW $>1200$~\AA} \label{sec:strong_eelg}

We have obtained high resolution spectra of $13$ galaxies with [O~{\small III}]+H$\beta$ EW $>1200$~\AA, and all these $13$ systems have systemic redshifts necessary for full characterization of the line profiles. The galaxies in this subset are both lower in stellar mass (median $M_\star=6.5\times10^7\ M_\odot$) and larger in sSFR (median $=127$~Gyr$^{-1}$) than those with [O~{\small III}]+H$\beta$ EW $=400-1200$~\AA. While our sample is small, the galaxies in this subset have larger Ly$\alpha$ EWs than seen in the rest of the sample (Figure~\ref{fig:ssfr_mass_lyaew_o3hbew}). The line profiles are shifted toward lower velocity offsets (median $v_{\rm peak}=82$~km~s$^{-1}$, with range $-18$ to $593$~km~s$^{-1}$) and larger central flux fractions (median $f_{{\rm cen,Ly}\alpha}=0.24$) (see Figure~\ref{fig:vpeak_fcen_o3hbew}). Five of the six galaxies with $f_{{\rm cen,Ly}\alpha}>0.38$ (as discussed in Section~\ref{sec:high_fcen}) are in the small subsample with [O~{\small III}]+H$\beta$ EW $>1200$~\AA. Based on these results, it does seem that the low H~{\small I} density channels (that are required for line center transmission of Ly$\alpha$) are more common among the galaxies with the very largest [O~{\small III}]+H$\beta$ EWs. 

However there are also $8$ galaxies with extremely strong line emission ([O~{\small III}]+H$\beta$ EW $>1200$~\AA) and relatively weak Ly$\alpha$ ($6-40$~\AA) with negligible emission at line center (median $f_{{\rm cen,Ly}\alpha}=0.1$) and large velocity offsets (median $v_{\rm peak}=230$~km~s$^{-1}$), both suggesting complete coverage of the H~{\small II} regions with reasonably dense neutral gas \citep[e.g.,][]{Erb2014,Hashimoto2015,Verhamme2015}. It is conceivable that the weaker Ly$\alpha$ emitters have yet to create the low density channels necessary for direct escape of Ly$\alpha$, perhaps reflecting an earlier evolutionary stage before feedback has disrupted the H~{\small I} \citep[e.g.,][]{Ma2015,Trebitsch2017,Barrow2020}. However there is no evidence from the SEDs that these $8$ systems (median light-weighted age $\simeq10$~Myr) are younger than the galaxies with Ly$\alpha$ escaping at line center (median light-weighted age $\simeq4$~Myr). Alternatively it may just be that some sightlines to galaxies with large sSFR are more likely to be cleared of H~{\small I} than others. In this case, it is plausible that the dispersion in Ly$\alpha$ profiles at [O~{\small III}]+H$\beta$ EW $>1200$~\AA\ reflects viewing angle effects, with only a subset of sources having low density sightlines oriented toward us \citep[e.g.,][]{Gnedin2008,Cen2015,Fletcher2019,Smith2019,Katz2020,Nakajima2020}. While such variations are likely to be present in individual galaxies, we have no observations that indicate that this is definitively the case for the systems in our sample. 


\begin{figure*}
\includegraphics[width=\linewidth]{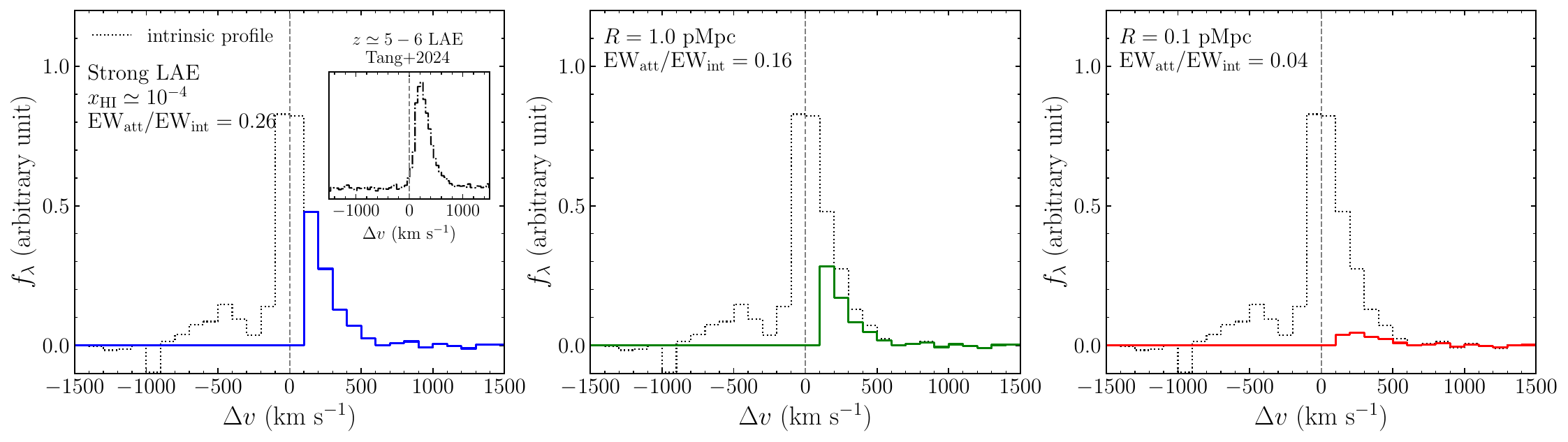}
\includegraphics[width=\linewidth]{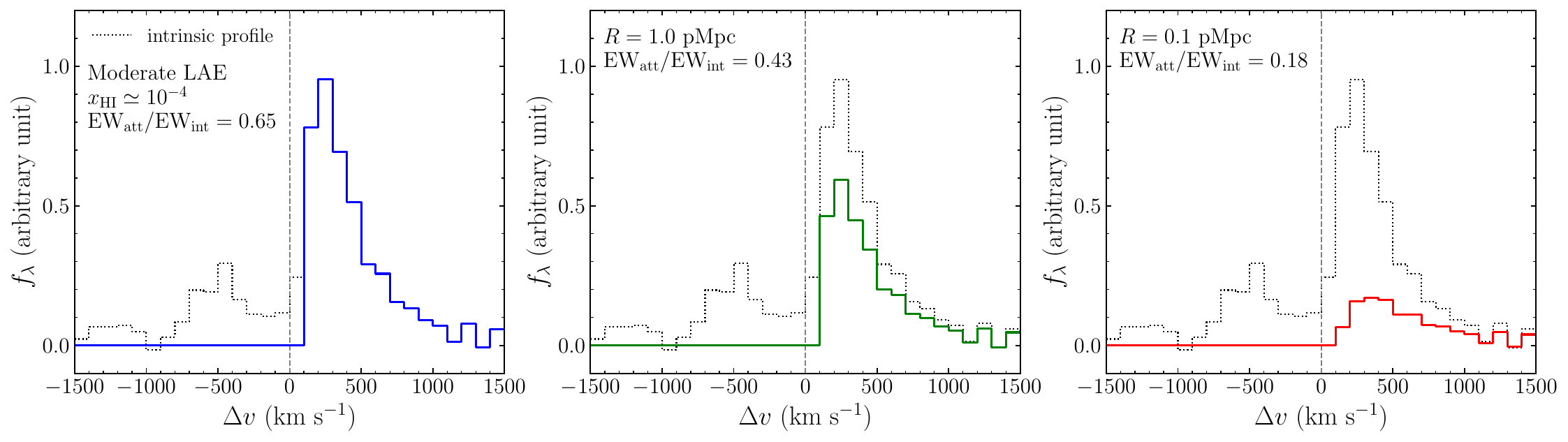}
\caption{Predicted impact of the IGM attenuation to Ly$\alpha$ profiles of galaxies at $z\gtrsim6$ (solid color lines). We assume the composite Ly$\alpha$ profile of strong Ly$\alpha$ emitters (EW $=80$~\AA) of $z=2.1-3.4$ EELGs as the ``intrinsic'' profile emerging from the ISM and the CGM (black dotted lines) in top panels and moderate Ly$\alpha$ emitters (EW $=20$~\AA) in bottom panels. In order to compare with {\it JWST}/NIRSpec results, we convolve the profiles with the resolution of NIRSpec grating ($R=1000$). In left panels, we consider the resonant scattering by the residual H~{\scriptsize I} in the IGM at $z=6$, which could attenuate the Ly$\alpha$ blueward the line center and the infalling IGM further scatters the Ly$\alpha$ redward. We overplot the composite Ly$\alpha$ profile of $z\simeq5-6$ galaxies \citep{Tang2024} as a comparison. In middle and right panels, we in addition consider the IGM damping wing absorption at $z=7$. We assume the galaxy is in an ionized bubble, sitting a distance $R=1.0$~pMpc (middle panels) or $R=0.1$~pMpc (right panels) from the neutral IGM. In each panel we list the fraction of Ly$\alpha$ photons transmitted through the IGM comparing to the intrinsic value (EW$_{\rm att}/{\rm EW}_{\rm int}$).}
\label{fig:lya_wing}
\end{figure*}

To better illustrate the physical factors regulating the Ly$\alpha$ profile in our sample, we plot $v_{\rm peak}$ vs. sSFR and $f_{{\rm cen,Ly}\alpha}$ vs. sSFR in Figure~\ref{fig:vpeak_fcen_ssfr}. At large sSFR ($\gtrsim100$~Gyr$^{-1}$; assuming CSFH) we continue to see high peak velocity offset ($v_{\rm peak}>200$~km~s$^{-1}$) sources with low central escape factions ($f_{{\rm cen,Ly}\alpha}<0.1$). Clearly a recent burst is likely necessary, but not a sufficient criteria for low density channels that facilitate direct Ly$\alpha$ escape. But in our sample, it is only in this high sSFR sample where we see the low peak velocity offsets ($v_{\rm peak}<100$~km~s$^{-1}$) and high central escape fraction ($f_{{\rm cen,Ly}\alpha}>0.2$) galaxies, indicating lower density sightlines, some of which may be conducive to LyC leakage (e.g., \citealt{Verhamme2015,Dijkstra2016b}, though with scatter; \citealt{Pahl2024}). This is consistent with the picture that in high sSFR galaxies, the strong stellar feedback associated with intense bursts can efficiently disrupt the neutral gas surrounding massive stars \citep[e.g.,][]{Kimm2019,Ma2020,Kakiichi2021}. We note that these systems also have low masses, which could imply reduced the H~{\small I} and dust as well \citep[e.g.,][]{Erb2014}. Detailed investigations of LyC leakers suggests low density sightlines may be more common in galaxies with low stellar mass \citep[e.g.,][]{Fletcher2019,Chisholm2022,Saldana-Lopez2023,Pahl2023}, whereas trends with sSFR are not as clear \citep{Pahl2023}. It is difficult to distinguish which effect (low mass or large sSFR) is responsible for the low peak velocity offset. It may be that it is the combination of the two factors, i.e. presence of a burst in a low mass galaxy which creates conditions optimal for low peak velocity offset. Larger samples with low masses spanning larger range of sSFR are required for more insight.

In the following section, we will discuss a sample of $z\gtrsim 7$ galaxies with Ly$\alpha$ emerging at very high velocities ($>500$~km~s$^{-1}$; \citealt{Bunker2023a,Jung2023,Tang2023}). We find one source in this subset of our sample with a highly-redshifted Ly$\alpha$ profile (UDS-08078), similar to the $z\gtrsim7$ galaxies. In spite of its very large [O~{\small III}]+H$\beta$ EW ($1321$~\AA), the Ly$\alpha$ escapes with a large peak velocity offset of $593$~km~s$^{-1}$ and a wide FWHM of $414$~km~s$^{-1}$. We find that $45\%$ of the line flux is redshifted to $>600$~km~s$^{-1}$. The $z\gtrsim7$ galaxies with large velocity offsets tend to be very luminous. UDS-08078 is similar, with an absolute magnitude (M$_{\rm UV}=-21.7$) that is significantly brighter than the median value in our sample. It is plausible that the most UV luminous galaxies have larger H~{\small I} column densities, shifting the emergent Ly$\alpha$ profile to higher velocities \citep[e.g.,][]{Mason2018b,Endsley2022b}. This would contribute to the visibility of Ly$\alpha$ in the most luminous galaxies at $z\gtrsim7$. Larger samples of luminous (M$_{\rm UV}\lesssim-21.5$) EELGs at lower redshifts (where the IGM is highly ionized) are required to determine whether the Ly$\alpha$ velocity offsets are uniformly high in this population.


\section{L\lowercase{y}$\alpha$ Profiles in the Reionization Era} \label{sec:discussion}

\subsection{Expectations for Ly$\alpha$ Profiles of EELGs at $z>6$} \label{sec:lya_model}

The Ly$\alpha$ line profiles of $z\gtrsim6$ galaxies are strongly impacted by the IGM, providing a sensitive measure of the local progress of reionization \citep[e.g.,][]{Mason2020,Endsley2022b,Saxena2023,Tang2024}. The utility of line profiles as a probe of the IGM relies on knowledge of the line shape before it is modulated by the IGM. Given their similarity to reionization-era galaxies, the Ly$\alpha$ spectra of $z\simeq2-3$ EELGs provide a useful baseline for understanding the line profiles now being observed at $z\gtrsim7$. We have found that the Ly$\alpha$ profiles of $z\simeq2-3$ EELGs (see Section~\ref{sec:OA}) appear very different from their counterparts at $z\gtrsim5$ (\citealt{Tang2024}; see also Section~\ref{sec:zg6_lya}), with small Ly$\alpha$ peak velocity offsets and often very large central flux fractions. In this subsection, we investigate whether this evolution follows naturally from the impact of the IGM on the centrally-peaked EELGs. We will assume that our database of $z\simeq2-3$ EELG spectra approximates the intrinsic Ly$\alpha$ profile, which we define as the profile emerging from the ISM and CGM. We will first consider the impact of the dense highly-ionized IGM at $z\simeq5-6$, then consider the additional impact of the damping wing as the IGM becomes more neutral at $z\gtrsim6$. 

To explore how the $z\simeq2-3$ Ly$\alpha$ profiles may appear in the reionization era, we first create a set of composites using the individual spectra obtained in this paper. We include galaxies in our sample with EW$_{{\rm [OIII]+H}\beta}=600-3000$~\AA. This range covers $\simeq65\%$ of the [O~{\small III}]+H$\beta$ EWs spanned by $z\simeq6.5-8$ galaxies \citep{Endsley2023}, and importantly spans the [O~{\small III}]+H$\beta$ EWs of most Ly$\alpha$ emitting galaxies at $z>7$ \citep[e.g.,][]{Schenker2014,Finkelstein2013,Zitrin2015,Oesch2015,Roberts-Borsani2016,Stark2017,Endsley2021a,Endsley2021b,Larson2022,Bunker2023a,Saxena2023,Tang2023}. We group galaxies by Ly$\alpha$ EW, creating one stack for the strongest Ly$\alpha$ emitters (EW$_{{\rm Ly}\alpha}=80$~\AA) and one for moderate Ly$\alpha$ emitters (EW$_{{\rm Ly}\alpha}=20$~\AA). To generate the composites, we shift individual spectra to the rest-frame using the systemic redshifts. Each spectrum is then interpolated to a common wavelength scale with a bin size ($0.125$~\AA\ in rest-frame) that is larger than the wavelength bin size of individual spectra. We then normalize each individual spectrum using its measured Ly$\alpha$ flux. Finally, the spectra are stacked by median-combining the individual flux densities in each wavelength bin. The composite spectra are shown in Figure~\ref{fig:lya_wing} as dotted lines, where the rest-frame wavelength is converted to the velocity space. As expected based on the individual profiles (Figure~\ref{fig:lya_spec1}), the stack of the strongest Ly$\alpha$ emitters in our sample has a very large ($f_{{\rm cen,Ly}\alpha}=0.47$) with Ly$\alpha$ peaking near line center. The stack of the more moderate Ly$\alpha$ emitters has a line profile with less emission near systemic ($f_{{\rm cen,Ly}\alpha}=0.08$) and the peak velocity occurring at $v_{\rm peak}=236$~km~s$^{-1}$. We now investigate how these two Ly$\alpha$ profiles would appear at higher redshifts where the IGM is considerably denser and more neutral.

We first consider the impact of the IGM on the composite Ly$\alpha$ profiles at $z\simeq5-6$. While the IGM is mostly ionized at these redshifts (albeit with non-negligible neutral fractions at $z\simeq 6$), the IGM density is large enough at $z\simeq 5$ for the residual neutral hydrogen ($x_{\rm HI}\gtrsim10^{-5}-10^{-4}$; e.g., \citealt{Yang2020b,Bosman2022}) to resonantly scatter the blue side of the line \citep{Gunn1965}. Given the line profiles we have presented in Section~\ref{sec:result}, it is clear this will have a significant impact on the recovered fluxes at the tail end of reionization. Considering the two $z\simeq2-3$ EELG composites described above, we find that the blue side of the line contains $45$ and $24\%$ of the total line flux for the strong and moderate Ly$\alpha$ stacks, respectively. This includes emission in the blue peak as well as in the blue half of the central flux component (see Figure~\ref{fig:lya_spec1} and \ref{fig:lya_spec2}). 
This indicates that the Ly$\alpha$ emission emitted by some EELGs may decrease in EW by up to a factor of two due to IGM attenuation between $z\simeq2-3$ and $z\simeq5-6$. 
At $z\simeq5-6$, the velocity profiles of EELGs should look distinctly different (i.e., sharper blue cutoff) than those of galaxies at $z\simeq2-3$ with similar Ly$\alpha$ EWs ($\simeq100$~\AA). It has been seen that Ly$\alpha$ emitters at $z\simeq5-6$ generally show negligible flux blueward the line center \citep{Tang2024}, consistent with the strong attenuation to the blue Ly$\alpha$ photons at $z\simeq5-6$.

Gravitational infall of gas from the IGM onto galaxies is predicted to further alter the line profiles at $z\simeq5-6$ \citep[e.g.,][]{Santos2004,Dijkstra2007,Laursen2011,Mason2018a}, resonantly scattering Ly$\alpha$ photons on the red side of the systemic redshift. This will not only decrease the Ly$\alpha$ EW relative to the $z\simeq2-3$, but it will shift the peak velocity to the red. To consider the impact of infalling IGM on the $z\simeq2-3$ EELG line profiles, we adopt the model used in \citet{Mason2018a}. Here gas is assumed to be infalling at the circular velocity of the halo. Halo masses are estimated from redshift and M$_{\rm UV}$ using the abundance matching relations presented in \citet{Mason2015}. While more detailed treatments of IGM infall are possible \citep[e.g.,][]{Santos2004,Dijkstra2007,Sadoun2017,Weinberger2018,Park2021}, these are beyond the scope required for the goals in this paper. In the \citet{Mason2018a} model, the impact of infall on the profile will tend to be greater in more luminous galaxies, as the larger halo masses will enable scattering further on the red side of the line. We apply this infall model to our two composite spectra. We adopt an absolute magnitude similar to those of faint galaxies (M$_{\rm UV}=-18$, corresponding to an infall velocity of $\simeq110$~km~s$^{-1}$) now being observed with Ly$\alpha$ at $z\gtrsim 5$ with {\it JWST}, but we will comment on how our results would change if we considered more luminous galaxies.


\begin{deluxetable*}{cccccccccc}
\tablecaption{Ly$\alpha$ peak velocity offsets of the $11$ galaxies at $z>6.5$ derived from public {\it JWST}/NIRSpec grating ($R\sim1000$) spectra. The systemic redshifts ($z_{\rm sys}$) are derived by fitting strong rest-frame optical emission lines (H$\beta$, [O~{\scriptsize III}]~$4959$, [O~{\scriptsize III}]~$5007$, or H$\alpha$) with Gaussian profiles. The Ly$\alpha$ redshifts ($z_{{\rm Ly}\alpha}$) are derived from the peak of Ly$\alpha$ emission lines.}
\tabletypesize{\small}
\tablehead{
Program & PID & NIRSpec ID & R.A. & Decl. & $z_{\rm sys}$ & $z_{{\rm Ly}\alpha}$ & M$_{\rm UV}$ & $v_{\rm peak}$ & Ref. \\
 & & & (hh:mm:ss) & (dd:mm:ss) & & & (mag) & (km s$^{-1}$) & 
}
\startdata
JADES & 1181 & 1899 & 12:36:47.46 & 62:15:25.10 & $8.2791$ & $8.2801$ & $-19.42$ & $32\pm47$ & [1], This work \\ 
JADES & 1181 & 1129 & 12:36:43.16 & 62:16:56.68 & $7.0866$ & $7.0899$ & $-19.44$ & $122\pm53$ & This work \\ 
JADES & 3215 & 20213084 & 03:32:38.14 & -27:45:54.25 & $8.4858$ & $8.4907$ & $-19.47$ & $156\pm82$ & [1],[2] \\ 
JADES & 1181 & 38420 & 12:36:42.04 & 62:16:56.15 & $6.7332$ & $6.7378$ & $-20.59$ & $178\pm56$ & This work \\ 
JADES & 1210 & 13682 & 03:32:40.20 & -27:46:19.12 & $7.2754$ & $7.2833$ & $-17.60$ & $217\pm94$ & [3],[4] \\ 
JADES & 1181 & 1075 & 12:36:48.63 & 62:16:31.83 & $6.9080$ & $6.9140$ & $-19.84$ & $228\pm55$ & This work \\ 
CEERS & 1345 & 1019 & 14:20:08.49 & 52:53:26.38 & $8.6784$ & $8.6877$ & $-22.09$ & $288\pm161$ & [5] \\ 
CEERS & 1345 & 1027 & 14:19:31.92 & 52:50:25.50 & $7.8188$ & $7.8280$ & $-20.73$ & $313\pm88$ & [5] \\ 
JADES & 1180 & 8532 & 03:32:34.93 & -27:47:01.85 & $6.8778$ & $6.8866$ & $-19.86$ & $335\pm164$ & This work \\ 
JADES & 1210 & 9903 & 03:32:40.56 & -27:46:43.65 & $6.6310$ & $6.6406$ & $-18.63$ & $377\pm102$ & [4], This work \\ 
CEERS & 1345 & 698 & 14:20:12.08 & 53:00:26.79 & $7.4703$ & $7.4854$ & $-21.70$ & $535\pm92$ & [5] \\ 
\enddata
\tablecomments{PID: {\it JWST} program ID. Ref.: [1] \citet{Witstok2024}; [2]. \citet{Tang2024}; [3]. \citet{Saxena2023}; [4]. \citet{Jones2024}; [5]. \citet{Tang2023}.}
\label{tab:zg6p5_lya}
\end{deluxetable*}

The results are shown in the top left panel of Figure~\ref{fig:lya_wing} for our strong centrally-peaked Ly$\alpha$ composite. As expected, the ionized IGM has a significant effect on the profile given the large fraction of flux near the systemic redshift. For a galaxy with M$_{\rm UV}=-18$, we see that the peak velocity shifts from line center to $140$~km~s$^{-1}$ with only $26\%$ of the line transmitted through the IGM. This suggests that the centrally-peaked Ly$\alpha$ emitters (which appear commonly in EELG samples at $z\simeq2-3$) should be somewhat rare at $z\simeq5-6$ owing to the resonant scattering by the neutral gas in the ionized IGM. This is exactly what is seen (top left panel of Figure~\ref{fig:lya_wing}; see also \citealt{Tang2024}). The more moderate Ly$\alpha$ emitter stack has a larger peak velocity and hence is less impacted by IGM infall (see the bottom left panel of Figure~\ref{fig:lya_wing}). Again considering a galaxy with M$_{\rm UV}=-18$, we see that the infall prescription results in a sharper blue cutoff, but the peak velocity ($250$~km~s$^{-1}$) is not significantly different from the assumed intrinsic profile. Owing to the larger peak velocity of the intrinsic profile, the transmission is less affected by the IGM, with $65\%$ of the line luminosity emerging. As a result, we expect Ly$\alpha$ emitter samples to be increasingly dominated by intrinsic profiles with larger peak velocities at $z\simeq 5-6$.

The Ly$\alpha$ profiles will be further altered at $z\gtrsim7$, as more of the IGM becomes neutral and the IGM damping wing begins to attenuate the red side of the line in typical galaxies \citep[e.g.,][]{Mason2018a,Hoag2019,Bolan2022,Nakane2023,Umeda2023}. To estimate the impact on the $z\simeq2-3$ Ly$\alpha$ profiles, we apply the damping wing optical depth of Ly$\alpha$ as a function of velocity offset from systemic \citep{Miralda-Escude1998,Dijkstra2016a}, while also applying the attenuation due to resonant scattering from the infalling IGM as described above. We consider a galaxy at $z=7$ situated at the center of an ionized bubble with a distance $R=1.0$ and $0.1$~pMpc from the neutral IGM. For simplicity, as the distance to the first neutral patch of gas dominates the damping wing optical depth \citep[e.g.,][]{Mesinger2008}, we assume the IGM is completely neutral outside the bubble. We again assume a UV-faint galaxy (M$_{\rm UV}=-18$), which minimizes the effect that resonant scattering from infalling IGM gas is likely to have on the lines in the \citet{Mason2018a} models, allowing the effect of the damping wing to be more clearly identified. 


\begin{figure*}
\includegraphics[width=\linewidth]{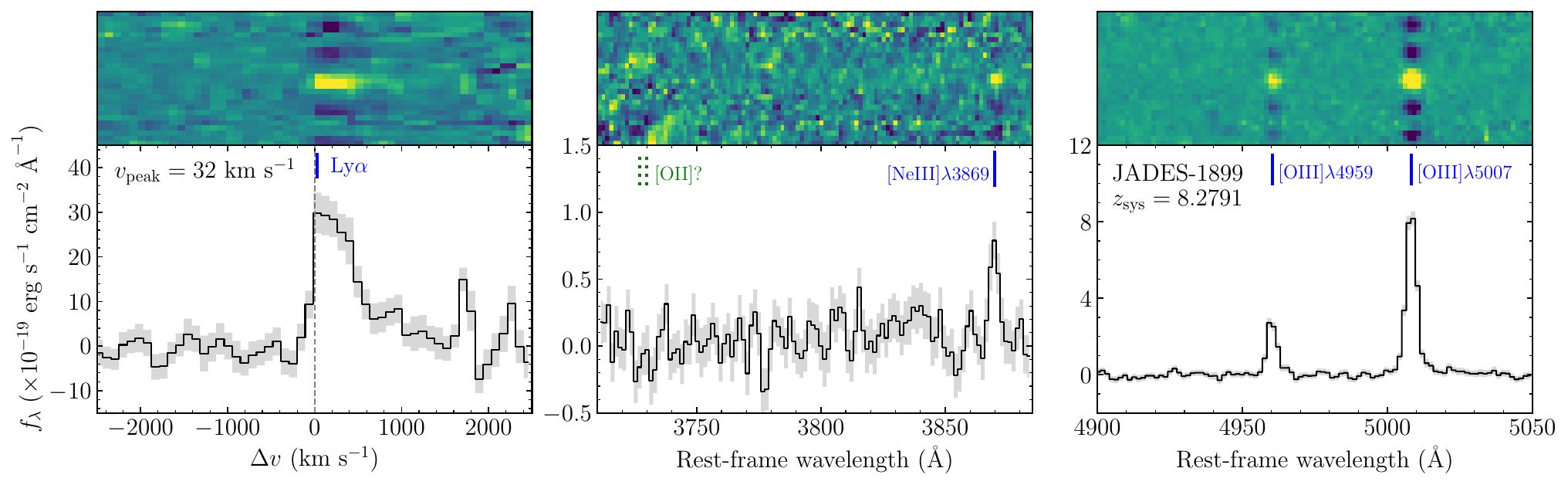}
\includegraphics[width=\linewidth]{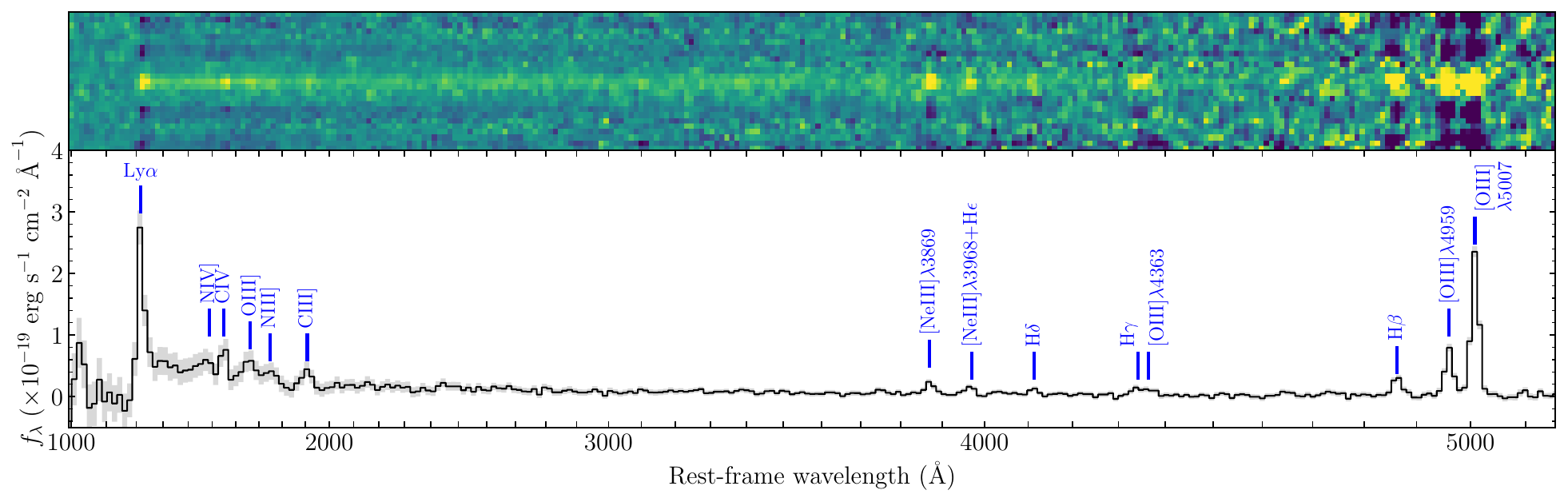}
\caption{{\it JWST}/NIRSpec 2D and 1D medium resolution ($R\sim1000$) grating spectrum (top) and low resolution ($R\sim100$) prism spectrum (bottom) of the strong Ly$\alpha$ emitting galaxy at $z=8.28$ (NIRSpec ID 1899) identified from the public JADES program 1181 dataset. More details of this galaxy are presented in \citet{Witstok2024}. The top left panel shows the Ly$\alpha$ velocity profile extracted from the G140M/F070LP spectrum. The Ly$\alpha$ peak is close to the systemic redshift (grey dashed line), with a peak velocity offset $v_{\rm peak}=32\pm47$~km~s$^{-1}$. The systemic redshift ($z_{\rm sys}=8.2791$) used to extract velocity profile is derived by fitting strong [O~{\scriptsize III}]~$\lambda5007$ and $4959$ emission lines from the G395M/F290LP spectrum (top right). We also find a clear [Ne~{\scriptsize III}]~$\lambda3869$ detection but we do not detect [O~{\scriptsize II}]~$\lambda\lambda3727,3729$ (top middle), suggesting a very large Ne3O2 ratio ($>2.6$ at $3\sigma$). The prism spectrum (bottom) shows various high ionization UV emission lines (N~{\scriptsize IV}], C~{\scriptsize IV}, O~{\scriptsize III}], N~{\scriptsize III}], and C~{\scriptsize III}]), indicating a hard ionizing spectrum of this galaxy.}
\label{fig:z8p3_lae_spec}
\end{figure*}

The resulting line profiles are shown in the top middle and the top right panels of Figure~\ref{fig:lya_wing} for the strong Ly$\alpha$ emitter composite. The impact of the IGM damping wing is very pronounced owing to the centrally-peaked profile of this composite. For the smallest bubbles considered ($R=0.1$~pMpc), these simple assumptions suggest the damping wing will only transmit a small fraction ($\simeq4\%$) of the line, converting an $80$~\AA\ Ly$\alpha$ emitter into a weak $3$~\AA\ detection, with the majority of line emission coming out at $100-500$~km~s$^{-1}$. For larger ionized regions, the IGM attenuation is still significant, with only $16\%$ of the line emerging in $R=1.0$~pMpc bubbles. We show the impact of the damping wing on the more moderate Ly$\alpha$ emitter composite in the bottom middle and the bottom right panels of Figure~\ref{fig:lya_wing}. These intrinsic profiles have larger peak velocities and thus face less attenuation, with IGM transmission ranging between $18\%$ to $43\%$ for $R=0.1$ and $1.0$ pMpc bubbles. Such systems should be visible deep in the reionization era, particularly in large ionized regions. 

\subsection{Ly$\alpha$ Velocity Offsets at $z>6$ with {\it JWST}} \label{sec:zg6_lya}


\begin{figure}
\includegraphics[width=\linewidth]{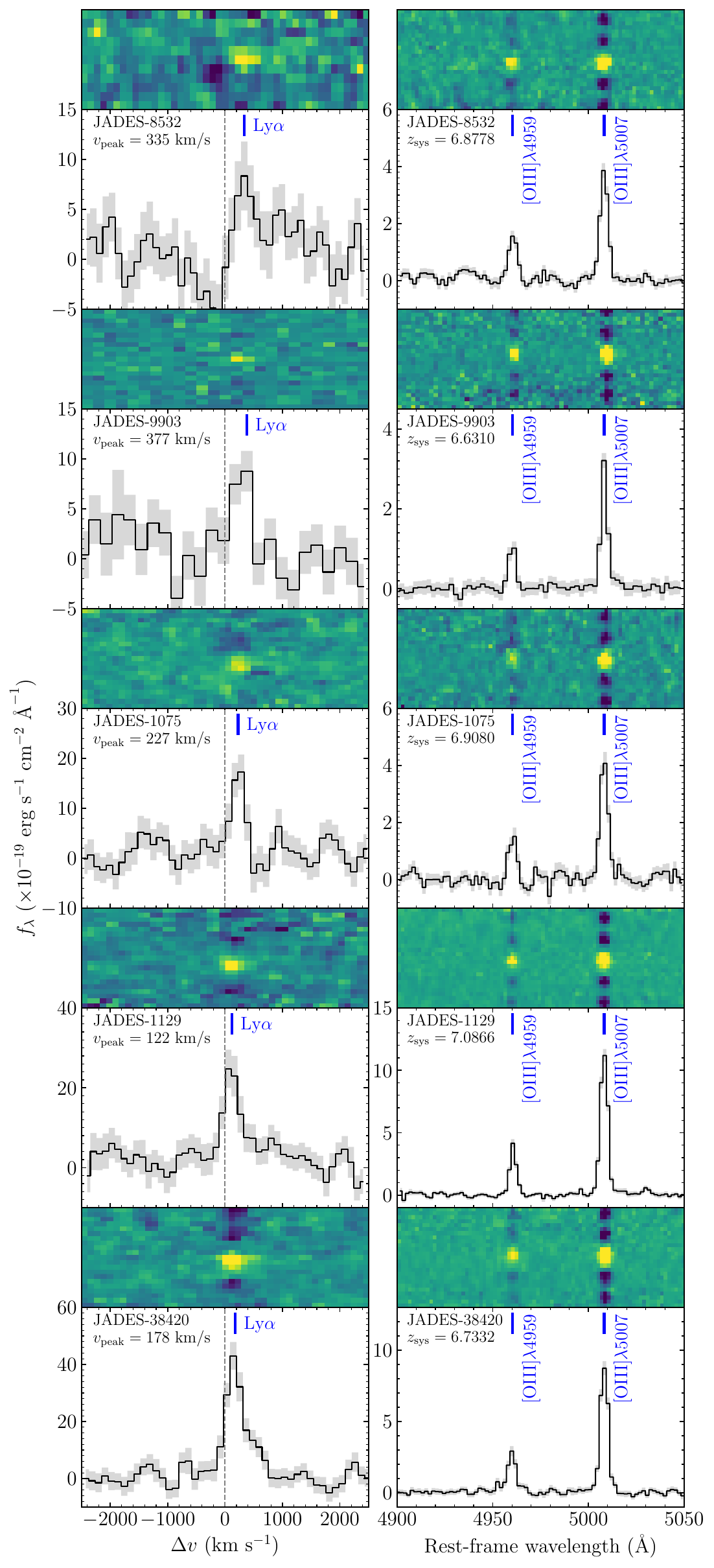}
\caption{2D and 1D {\it JWST}/NIRSpec medium resolution ($R\sim1000$) grating spectra of the five newly discovered Ly$\alpha$ emitting galaxies at $z>6.5$ from the public NIRSpec dataset (see Table~\ref{tab:zg6p5_lya}, except JADES-1899 shown in Figure~\ref{fig:z8p3_lae_spec}). For each object we show the Ly$\alpha$ velocity profile in the left and the [O~{\small III}]~$\lambda4959$ and $5007$ detections in the right. The velocity spaces are converted from the wavelength spaces using the systemic redshifts derived by fitting strong rest-frame optical emission lines with Gaussian profiles.}
\label{fig:zg6p5_lya}
\end{figure}

In Section~\ref{sec:lya_model}, we demonstrated that we expect significant evolution in the Ly$\alpha$ profiles of EELGs between $z\simeq2-3$ and $z\simeq5-6$, with galaxies having centrally-peaked Ly$\alpha$ profiles mostly disappearing from Ly$\alpha$-selected samples at $z\gtrsim5$. This has now been shown to occur in \citet{Tang2024}, leveraging systemic redshifts from the NIRCam grism in fields with ground-based Ly$\alpha$ detections. In this subsection, we extend the work in \citet{Tang2024} to Ly$\alpha$ peak velocities at $z\gtrsim6.5$, redshifts where the damping wing may have a stronger effect on the line profiles of sources with detectable Ly$\alpha$ emission. Here we focus on galaxies with both Ly$\alpha$ detections and systemic redshifts derived from NIRSpec, considering only those systems observed with the medium or high resolution gratings, as the prism does not give adequate velocity resolution to measure robust line profiles. We are in particular interested in whether galaxies with small peak velocities and large escape fractions can be identified and linked to large ionized bubbles. 

Our sample is selected from our own reductions as part of an ongoing effort to build a complete database of Ly$\alpha$ measurements in the reionization era. 
Our focus in this paper is on galaxies spectroscopically confirmed at $z>6.5$ using the public {\it JWST}/NIRSpec medium resolution ($R\sim1000$) or high resolution ($R\sim2700$) spectra dataset. 
Full details of this study will be described in a future paper (Tang et al. in prep.), but we will present the full sample with peak velocities in this paper.
In brief, we take grating spectra from the following public NIRSpec observations: 
the {\it JWST} Advanced Deep Extragalactic Survey (JADES; \citealt{Bunker2023b,Eisenstein2023a,Eisenstein2023b}), the GLASS-{\it JWST} Early Release Science Program \citep{Treu2022}, and the Cosmic Evolution Early Release Science (CEERS; \citealt{Finkelstein2023}). 
The JADES NIRSpec observations were performed with the low spectral resolution ($R\sim100$) PRISM/CLEAR setup and the medium resolution ($R\sim1000$) G140M/F070LP and G395M/F290LP grating/filter setups. 
The GLASS NIRSpec observations were performed with the high resolution ($R\sim2700$) G140H/F100LP, G235H/F170LP, and G395H/F290LP grating/filter setups. 
The CEERS NIRSpec observations were performed with the PRISM/CLEAR and the G140M/F100LP, G235M/F170LP, and G395M/F290LP setups. 
The NIRSpec spectra used here were reduced by one of the co-authors (M. Topping) following the procedures described in \citet{Tang2023}. 

From the above dataset, we have identified $75$ galaxies at $z>6.5$ based on detections of rest-frame optical emission lines in medium or high resolution NIRSpec grating spectra. 
For these sources, we derive the systemic redshifts by fitting Gaussians to the available strong optical emission lines (H$\beta$, [O~{\small III}], or H$\alpha$), as in \citet{Tang2023}.
We then search for Ly$\alpha$ emission lines using the systemic redshifts. 
Among the $56$ galaxies at $z>6.5$ with NIRSpec grating spectra, we have identified Ly$\alpha$ emission lines in $11$ objects. 
For each object with Ly$\alpha$ emission, we derive the Ly$\alpha$ redshift by fitting the Ly$\alpha$ line with a Gaussian (e.g., JADES-1075; Figure~\ref{fig:zg6p5_lya}) or a truncated Gaussian (e.g., JADES-1899; Figure~\ref{fig:z8p3_lae_spec}) to account for for the impact of the IGM on the blue side of the line. 
With both Ly$\alpha$ and systemic redshifts, we calculate the Ly$\alpha$ peak velocity offset. 
To evaluate the uncertainty of Ly$\alpha$ peak velocity offset, we resample the flux densities $1000$ times by taking the observed flux densities as mean values and the errors as standard deviations. 
We measure the redshifts and derive the Ly$\alpha$ peak velocity offsets from the resampled spectra and take the standard deviation of $v_{\rm peak}$ as the uncertainty. 
The derived values of the peak offsets range from $32$~km~s$^{-1}$ to $535$~km~s$^{-1}$, with a median $v_{\rm peak}=228$~km~s$^{-1}$ (Table~\ref{tab:zg6p5_lya}). 
For five sources of the eleven galaxies, the NIRSpec spectra and Ly$\alpha$ detections have previously been reported in literature (CEERS-698, CEERS-1019, CEERS-1027; \citealt{Tang2023}; JADES-13682; \citealt{Saxena2023}; JADES-20213084; \citealt{Tang2024,Witstok2024}). 
In Figure~\ref{fig:z8p3_lae_spec} and Figure~\ref{fig:zg6p5_lya}, we show the Ly$\alpha$ velocity profiles of the other six newly-reported Ly$\alpha$ emitting galaxies at $z>6.5$ (we note that the galaxy shown in 
Figure~\ref{fig:z8p3_lae_spec} is also presented in \citealt{Witstok2024}). 


\begin{figure}
\includegraphics[width=\linewidth]{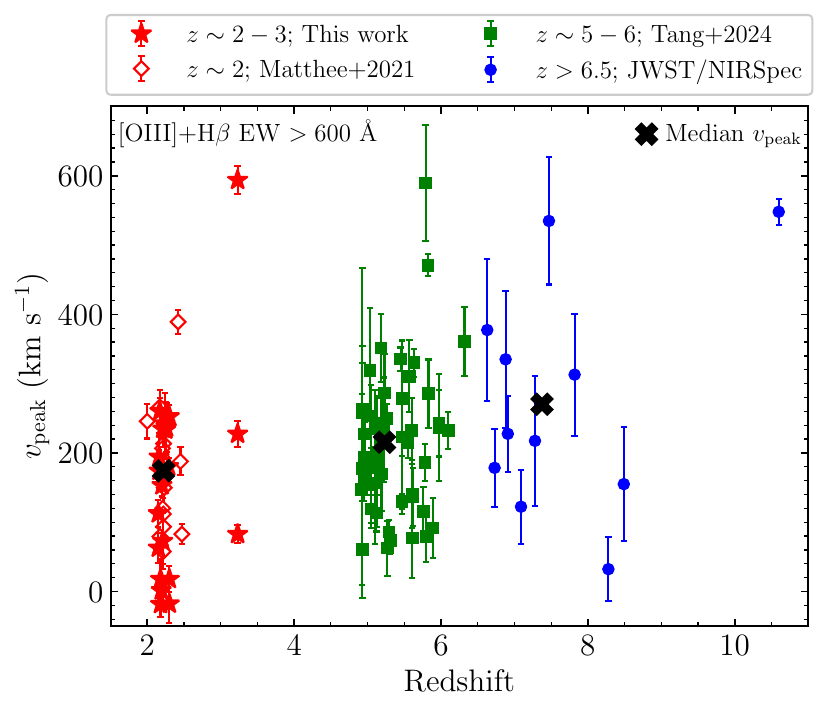}
\caption{Ly$\alpha$ peak velocity offset evolution for EELGs ([O~{\scriptsize III}]+H$\beta$ EW $>600$~\AA) over cosmic time. Here we limit to galaxies with M$_{\rm UV}>-22$. Our $z=2.1-3.4$ EELGs are shown by filled red stars. We overplot the $z\sim2$ XLS-$z2$ sample \citep{Matthee2021} with open red diamonds. Ly$\alpha$ emitters at $z\simeq5-6$ are presented by green squares \citep{Tang2024}. We also add Ly$\alpha$ peak velocity offsets of $z>6.5$ galaxies measured from the publicly available {\it JWST}/NIRSpec grating spectra (blue circles; Tang et al. in prep.). We show the median Ly$\alpha$ peak velocity offset of each subsample with black cross symbols.}
\label{fig:vpeak_z}
\end{figure}

We can now compare the distribution of Ly$\alpha$ profiles at $z\gtrsim6.5$ to those at $z\simeq2-3$.
In Figure~\ref{fig:vpeak_z} we show the Ly$\alpha$ peak velocity offsets as a function of redshift, limiting our sample to those with [O~{\small III}]+H$\beta$ EW $>600$~\AA.
For $z>6.5$ EELGs with M$_{\rm UV}>-22$, we find that Ly$\alpha$ peak velocity offsets are larger (median $v_{\rm peak}=230$~km~s$^{-1}$) than those seen in EELGs with strong Ly$\alpha$ emission at $z\simeq2-3$ (median $v_{\rm peak}=20$~km~s$^{-1}$). 
This result naturally follows our discussion in Section~\ref{sec:lya_model}, with the partially neutral IGM at $z>6.5$ preferentially attenuating the Ly$\alpha$ photons near line center (Figure~\ref{fig:lya_wing}). 
As has been described elsewhere \citep[e.g.,][]{Endsley2022b,Prieto-Lyon2023}, this will act to weaken Ly$\alpha$ in the galaxies with the smallest velocity offsets, likely shifting the distribution of peak velocities to the subset with Ly$\alpha$ centered at larger redshifts. 
In \citet{Tang2024}, we showed this evolution is already in place at $z\simeq 5-6$. 
Here we see a similar trend at yet higher redshifts where the damping wing will play a more prominent role. 

\subsection{An Intense Ly$\alpha$ Emitting Galaxy at $z>8$ with a Small Velocity Offset and Hard Radiation Field} \label{sec:z8p3_lae}

In \citet{Tang2024}, we presented a galaxy at $z=8.49$ with a relatively small velocity offset ($156$~km~s$^{-1}$) compared to most reionization-era galaxies. 
Here we find another $z\gtrsim8$ galaxy (JADES-1899 at $z=8.28$) with an even smaller peak velocity, potentially requiring very different surrounding IGM than most $z\gtrsim8$ galaxies discovered to date.
This galaxy is also recently reported in \citet{Witstok2024}. 
In the top of Figure~\ref{fig:z8p3_lae_spec}, we show the medium resolution grating spectrum of JADES-1899. 
Strong [O~{\small III}]~$\lambda4959$ and $5007$ emission lines are clearly seen in its NIRSpec G395M/F290LP spectrum. 
By fitting [O~{\small III}] with Gaussian profiles we derive a systemic redshift of $z_{\rm sys}=8.2791$. 
In the G140M/F070LP spectrum, we detect the Ly$\alpha$ emission line with $z_{{\rm Ly}\alpha}=8.2801$. 
This indicates a Ly$\alpha$ peak velocity close to the systemic redshift, with $v_{\rm peak}=32\pm47$~km~s$^{-1}$. 
The line profile is asymmetric, cutting off sharply at line center with minimal blue-sided emission. 
In the remainder of this subsection, we explore the nature of this source in more detail, with a goal of understanding how such a line profile can exist at $z\gtrsim8$.

In addition to its unique line profile, JADES-1899 also has one of the strongest Ly$\alpha$ lines yet reported at $z\gtrsim 8$ (see also \citealt{Fujimoto2023,Kokorev2023}). 
We measure a Ly$\alpha$ flux of $F_{{\rm Ly}\alpha}=7.30\pm0.52\times10^{-18}$~erg~s$^{-1}$~cm$^{-2}$ and EW $=137\pm10$~\AA. 
This is well above what has been typically seen to-date in $z\gtrsim 8$ star forming galaxies with {\it JWST} \citep{Bunker2023a,Tang2023,Tang2024,Jones2024,Saxena2024}, where typical sources have EWs that are $10\times$ weaker.
We also constrain the Ly$\alpha$ escape fraction using the H$\beta$ emission line. 
Because the S/N of the H$\beta$ detection in the grating spectrum is low ($\simeq3$), we calculate the Ly$\alpha$ escape fraction using the Ly$\alpha$ and H$\beta$ flux measured from the prism spectrum (bottom panel of Figure~\ref{fig:z8p3_lae_spec}) which has higher S/N ($\simeq6$). 
We first correct the H$\beta$ flux for dust attenuation using Balmer decrement measurement. 
The H$\gamma$/H$\beta$ ratio measured from prism spectrum is $0.474\pm0.129$. 
Comparing to the intrinsic H$\gamma$/H$\beta$ ratio expected in case B recombination ($0.468$ assuming an electron temperature $T_{\rm e}=10^4$~K; \citealt{Osterbrock2006}), this indicates negligible dust attenuation to the nebular emission. 
Then assuming case B recombination with $T_{\rm e}=10^4$~K and an electron density $n_{\rm e}=10^2$~cm$^{-3}$, we derive a Ly$\alpha$ escape fraction $f^{\rm case\ B}_{{\rm esc,Ly}\alpha}=0.34\pm0.06$. 
We note that if this galaxy leaks Ly$\alpha$ through optically-thin H~{\small I} gas, case A recombination may be a better approximation. 
Assuming case A recombination, the Ly$\alpha$ escape fraction is about $1.3$ times lower than the value derived from case B recombination, with $f^{\rm case\ A}_{{\rm esc,Ly}\alpha}=0.26\pm0.05$. 
JADES-1899 appears to be transmitting a much larger fraction of its Ly$\alpha$ than typical systems at $z\gtrsim8$. 
At these redshifts, galaxies with detectable Ly$\alpha$ emission are generally found to have very low escape fractions ($f^{\rm case\ B}_{{\rm esc,Ly}\alpha}=0.03-0.09$), consistent with significant damping wing attenuation from the IGM \citep[e.g.,][]{Bunker2023a,Tang2023}. 

Both the Ly$\alpha$ escape fraction and line profile in JADES-1899 point to reduced attenuation from the IGM. 
While this may suggest the galaxy resides in a long ionized sightline (as we will discuss below), it is important to note that this also requires minimal impact from the infalling IGM in the vicinity of the galaxy. 
As we demonstrated in Figure~\ref{fig:lya_wing}, Ly$\alpha$ profiles are expected to be mostly devoid of emission at $\Delta v=0-100$~km~s$^{-1}$ in cases where infall is important. 
This is observed in nearly all $z\gtrsim5$ galaxies (Figure~\ref{fig:vpeak_z}).
The presence of significant line flux at $\gtrsim30$~km~s$^{-1}$ in JADES-1899 requires that the red side of line center is not resonantly scattered on small scales by infalling gas. 
We will come back to discuss physical factors that may reduce the impact of infall in JADES-1899 at the end of this subsection.

\begin{figure}
\includegraphics[width=\linewidth]{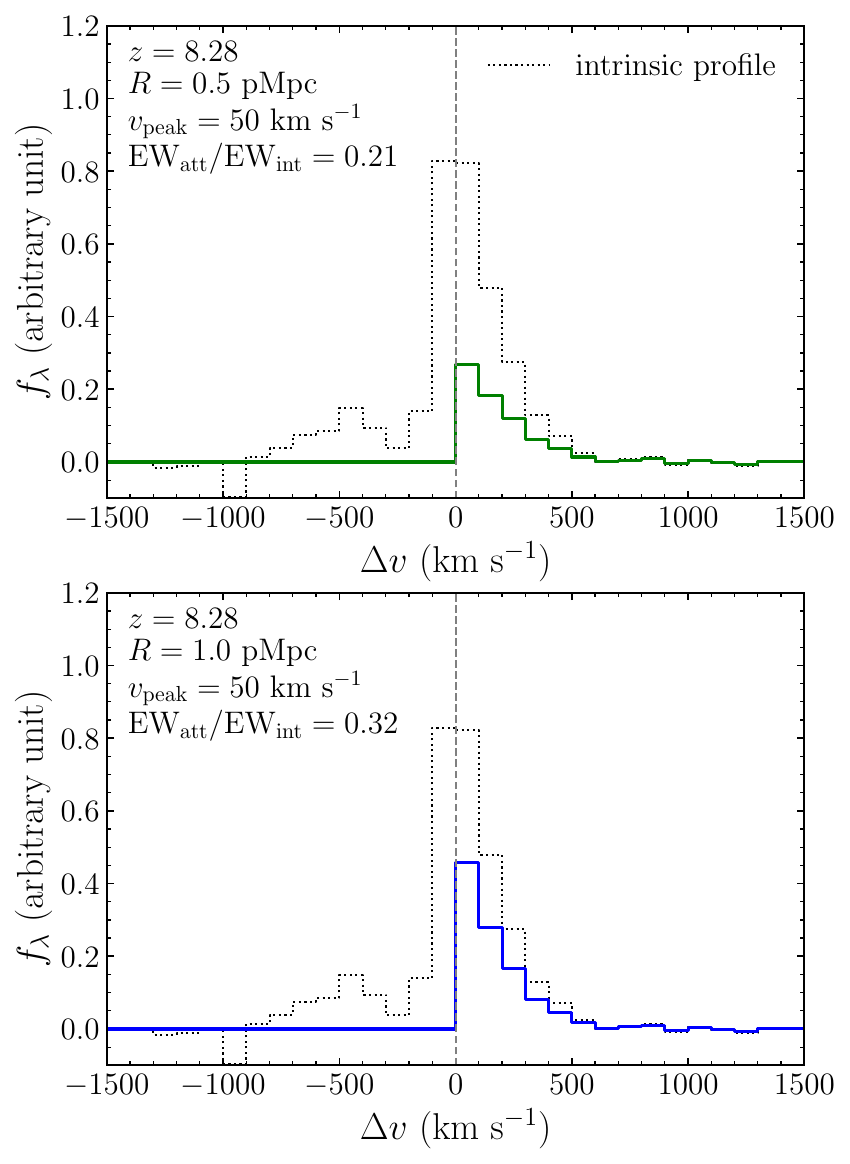}
\caption{Requirements for observing small Ly$\alpha$ velocity offsets at $z\gtrsim 8$. We assume the composite Ly$\alpha$ profile of strong Ly$\alpha$ emitters of $z=2.1-3.4$ EELGs as the ``intrinsic'' profile (black dotted lines). The spectra have been convolved with the resolution of NIRSpec ($R=1000$). In the top (bottom) panel, we consider the damping wing attenuation assuming the galaxy is centered in an ionized bubble sitting a distance $R=0.5$~pMpc ($R=1.0$~pMpc) from the neutral IGM. To recover the emission near line center, we must ignore the impact of IGM infall. The Ly$\alpha$ peak velocity is shifted to $v_{\rm peak}\simeq50$~km~s$^{-1}$, close to the $v_{\rm peak}$ seen in JADES-1899. The Ly$\alpha$ EW is attenuated by $\simeq3-5\times$ after IGM attenuation (EW$_{\rm att}/{\rm EW}_{\rm int}\simeq0.2-0.3$).}
\label{fig:lya_wing_z8p3}
\end{figure}

We also require the effects of the damping wing be small enough for us to recover $\simeq34\%$ of the Ly$\alpha$ luminosity, consistent with the recovered Ly$\alpha$ escape fraction. 
To roughly estimate the range of IGM environments that can reproduce the JADES-1899 Ly$\alpha$ profile, we again alter our $z\simeq2-3$ Ly$\alpha$ profiles following the same methodology applied in Section~\ref{sec:lya_model}. 
Given the small peak velocity of JADES-1899, we assume that the intrinsic profile is that of our strong Ly$\alpha$ composite shown in top panels of Figure~\ref{fig:lya_wing} (with significant emission at line center). 
We apply the damping wing optical depth of Ly$\alpha$ at $z=8.28$, assuming the galaxy is in a ionized bubble, sitting a distance $R=0.5$ or $1.0$~pMpc from neutral gas.
We assume that the small residual fraction of neutral hydrogen inside the ionized bubble resonantly scatters the Ly$\alpha$ emission blueward of line center, but we assume that the effects of IGM infall on the red side of the line are negligible. 
The results are shown in Figure~\ref{fig:lya_wing_z8p3}.
In a moderate ($R=0.5$~pMpc; top panel) or large ($R=1.0$~pMpc; bottom panel) bubble, we find that the Ly$\alpha$ peak velocities have shifted from line center to values that are consistent with the observed value of JADES-1899 within $1\sigma$ uncertainty ($v_{\rm peak}\simeq50$~km~s$^{-1}$). 
The transmission ranges between $21-32\%$ in these cases, consistent with the inferred Ly$\alpha$ escape fraction of JADES-1899. 
We note these calculations suggest that the Ly$\alpha$ profile of JADES-1899 can be explained with a smaller bubble than the $R\sim3$~pMpc bubble size predicted in the analysis of \citet{Witstok2024}.
While these calculations are mostly meant as illustrative of the possible range of bubbles that might host JADES-1899 ($R \gtrsim 0.5-1$ pMpc), it is clear that much smaller bubbles (e.g., $R=0.1$~pMpc) would result in stronger attenuation just redward of line center. 
In this case, the peak velocities would be shifted to larger values ($\simeq250$~km~s$^{-1}$) and would give an escape fraction ($\simeq4\%$) that is much smaller than observed. 


The IGM conditions that give rise to the Ly$\alpha$ profile of JADES-1899 must be rare at $z\gtrsim8$. 
On one hand this may suggest that the moderate-size bubbles ($\gtrsim0.5-1$ pMpc) required to explain the transmission are not common at $z\gtrsim8$, as would be expected at very early stages of reionization when neutral fractions are very large ($x_{\rm HI}\simeq0.9$; \citealt{Lu2024}). 
But the absence of significant attenuation from infalling IGM may also reflect unique conditions on smaller scales. 
For example, a hard radiation field could decrease the residual H~{\small I} fraction in the ionized IGM surrounding JADES-1899, reducing the impact of resonant scattering on the line profile \citep{Mason2020} or we may be observing Ly$\alpha$ escaping along a sightline without significant infall. Simulations predict a broad distribution of infall velocities whereby complex gas dynamics and strong outflows can counteract the spherical infall of IGM gas \citep[e.g.,][]{Iliev2008,Muratov2015,Park2021}. 
The spectrum of JADES-1899 is consistent with this picture, revealing a suite of intense rest-frame UV and optical emission lines seen in only a handful of early galaxies (Figure~\ref{fig:z8p3_lae_spec}).
In particular, we detect high ionization UV emission lines in the prism spectrum (Figure~\ref{fig:z8p3_lae_spec}). 
The most prominent line is C~{\small IV}~$\lambda\lambda1548,1551$, with unresolved doublet EW $=49\pm16$~\AA. 
Such large EW C~{\small IV} emission is extremely rare at lower redshifts \citep{Senchyna2017,Berg2019,Izotov2024} but is present in a small subset of reionization-era galaxies with metal poor gas \citep[e.g.,][]{Stark2015,Castellano2024,Topping2024}. 
We also find tentative detections (S/N $\simeq2$) of blended He~{\small II}~$\lambda1640$+O~{\small III}]~$\lambda\lambda1661,1666$ (EW $=65\pm25$~\AA), C~{\small III}]~$\lambda\lambda1907,1909$ (EW $=50\pm22$~\AA), and nitrogen emission lines N~{\small IV}]~$\lambda\lambda1483,1486$ (EW $=35\pm20$~\AA) and N~{\small III}]~$\lambda\lambda1746,1748$ (EW $=35\pm22$~\AA). 
The rest-frame optical presents a similar picture, with an intense [O~{\small III}] emission ([O~{\small III}]~$\lambda5007$ EW $=1625\pm113$~\AA) and an extremely large ionization-sensitive line ratio ([O~{\small III}]/[O~{\small II}] $>40$).
We may thus be observing JADES-1899 at a special time when its hard radiation field is facilitating enhanced transmission of Ly$\alpha$ near line center, while a relatively long ionized sightline is simultaneously reducing the impact of the damping wing on Ly$\alpha$.


\section{Summary} \label{sec:summary}

We have presented resolved ($R\simeq3900$) Ly$\alpha$ profiles of $42$ extreme [O~{\small III}] line emitting galaxies at $z=2.1-3.4$, which have [O~{\small III}]+H$\beta$ EWs ($\simeq300-3000$~\AA) that are similar to the range seen in reionization-era systems \citep[e.g.,][]{DeBarros2019,Endsley2021a,Endsley2023}. Twenty six of the forty two sources in our sample have systemic redshift measurements, enabling us to derive their detailed Ly$\alpha$ profile properties. We use these to investigate the neutral hydrogen distribution in the ISM and the CGM of galaxies with properties similar to those of the reionization era. Using our database of Ly$\alpha$ spectra, we consider how the IGM is likely to modify the Ly$\alpha$ profiles at $z\gtrsim6$. Below we summarize our findings.

1. We have identified six sources with extremely large Ly$\alpha$ central escape fraction ($f_{{\rm cen,Ly}\alpha}\gtrsim0.4$) in our sample, indicating that EELGs occasionally create very low density H~{\small I} channels allowing direct escape of Ly$\alpha$ photons. These galaxies have the largest [O~{\small III}]+H$\beta$ EWs (median $\simeq1900$~\AA) and the largest Ly$\alpha$ EWs (median $z\simeq90$~\AA) among objects in our sample. SED fitting indicates that the light of these six large $f_{{\rm cen,Ly}\alpha}$ systems is dominated by extremely young populations ($2-8$~Myr assuming CSFH) with very large sSFRs ($115-400$~Gyr$^{-1}$) and low stellar masses ($3\times10^7-1\times10^8\ M_{\odot}$), suggesting that low density H~{\small I} channels can form in low mass systems undergoing intense bursts of star formation. Those galaxies with large central flux fractions are often seen with a red tail of Ly$\alpha$ emission (extending to $300-400$~km~s$^{-1}$) and a blue peak, potentially indicating that the very low density H~{\small I} gas is surrounded by denser H~{\small I} which scatters Ly$\alpha$ to larger velocities.

2. Galaxies with [O~{\small III}]+H$\beta$ EWs ($=400-1200$~\AA) that are typical of the range seen in the reionization era (median EW $=780$~\AA; e.g., \citealt{Endsley2023}) present a range of Ly$\alpha$ profiles. Most exhibit moderate strength Ly$\alpha$ (median EW $=22$~\AA) with large Ly$\alpha$ peak velocity offsets (median $v_{\rm peak}=193$~km~s$^{-1}$) and small Ly$\alpha$ central escape fractions (median $f_{{\rm cen,Ly}\alpha}=0.1$), consistent with the standard expectations from backscattering and resonant scattering through the outflowing H~{\small I}. This indicates that EELGs often lack the low density H~{\small I} channels required for direct escape of Ly$\alpha$. Due to the dense H~{\small I} in the ISM or the CGM, Ly$\alpha$ flux in these systems is often scattered to very high velocities of $500-1000$~km~s$^{-1}$. About $21\%$ (on average) of the Ly$\alpha$ flux of [O~{\small III}]+H$\beta$ EW $=400-1200$~\AA\ galaxies in our sample is in blue peak Ly$\alpha$ with velocity $=-650$ to $-110$~km~s$^{-1}$.

3. Our sample contains $13$ galaxies with the largest [O~{\small III}]+H$\beta$ EWs ($>1200$~\AA). This population is rare at $z\simeq2-3$ but becomes more common at $z>6$. These systems are characterized by both low stellar masses (median $=6.5\times10^7\ M_{\odot}$) and very large sSFRs (median $=127$~Gyr$^{-1}$). Five of these thirteen galaxies present large $f_{{\rm cen,Ly}\alpha}$ ($\gtrsim0.4$) and low $v_{\rm peak}$ ($<100$~km~s$^{-1}$), indicating low density sightlines which may be conducive to LyC leakage. On the other hand, the other eight galaxies with [O~{\small III}]+H$\beta$ EW $>1200$~\AA\ in our sample show relatively weak Ly$\alpha$ EWs ($=6-40$~\AA) with larger $v_{\rm peak}$ (median $=230$~km~s$^{-1}$) and negligible Ly$\alpha$ flux near the line center (median $f_{{\rm cen,Ly}\alpha}=0.1$). These suggest more uniformly covered dense H~{\small I} gas surrounding H~{\small II} regions. The most UV-luminous (M$_{\rm UV}=-21.5$) galaxy in our sample shows the largest $v_{\rm peak}$ ($=593$~km~s$^{-1}$) with a wide FWHM ($=414$~km~s$^{-1}$). It is possible that such luminous galaxies have denser H~{\small I} columns that scatter Ly$\alpha$ to higher velocities, though Ly$\alpha$ spectroscopy of a larger sample of luminous EELGs (M$_{\rm UV}<-21.5$) at redshifts where the IGM is highly ionized is required to examine this scenario.

4. Assuming the Ly$\alpha$ profiles in our $z\simeq 2-3$ sample are similar to those emerging from $z\gtrsim 6$ galaxies, we explore how the profiles will be modulated by the IGM in the reionization era. 
At $z\simeq5-6$, the Ly$\alpha$ flux blueward the systemic redshift will likely be highly attenuated due to the residual H~{\small I} in high density IGM (i.e., the Gunn-Peterson effect). 
The infall of IGM at $z\simeq5-6$ could further scatter the Ly$\alpha$ redward the systemic redshift, shifting the Ly$\alpha$ peak to $140-250$~km~s$^{-1}$ in galaxies with M$_{\rm UV}=-18$. 
We also consider the impact of the neutral IGM on the $z\simeq2-3$ Ly$\alpha$ profiles at $z\gtrsim7$ in $R=0.1$ and $1.0$~pMpc ionized bubbles. 
The IGM damping wing together with the infall of IGM will highly attenuate the subset of EELGs with strong centrally-peaked Ly$\alpha$ emission, transmitting only $4$ ($R=0.1$~pMpc) to $16$~percent ($R=1.0$~pMpc) of the line radiation. 
In many EELGs in our sample, Ly$\alpha$ emerges with a larger peak velocity ($\simeq250-350$~km~s$^{-1}$), allowing a greater fraction ($18-43\%$) of the line to be transmitted through the damping wing of the neutral IGM. 
Our results suggest that the reionization-era IGM will lead to a disappearance of the centrally-peaked Ly$\alpha$ emitters which appear commonly in $z\simeq2-3$ EELG samples. This picture is consistent with $z>5$ Ly$\alpha$ detections \citep{Tang2024}. 

5. We present the Ly$\alpha$ peak velocity offsets of $11$ galaxies at $z>6.5$ derived from public {\it JWST}/NIRSpec dataset. 
At fixed M$_{\rm UV}$ and [O~{\small III}]+H$\beta$ EW, galaxies at $z>6.5$ show much larger Ly$\alpha$ peak offsets (median $v_{\rm peak}=230$~km~s$^{-1}$) than the strong Ly$\alpha$ emitters at $z\simeq2-3$ (median $v_{\rm peak}=20$~km~s$^{-1}$), reflecting the partially neutral IGM at $z>6.5$ preferentially attenuating the Ly$\alpha$ near the systemic redshift. 
We report a new Ly$\alpha$ emitter at $z=8.3$ identified from the JADES program 1181. 
This object is also presented in \citet{Witstok2024}. 
We measure its Ly$\alpha$ emission with a very large EW $=137$~\AA\ and a low peak velocity offset $v_{\rm peak}=32$~km~s$^{-1}$. 
The low peak velocity offset and the relatively large Ly$\alpha$ escape fraction ($f^{\rm case\ B}_{{\rm esc,Ly}\alpha}=0.34$) suggest that this galaxy is likely to be situated in a fairly long ($\gtrsim0.5-1.0$~pMpc) ionized sightline at $z=8.3$.
Detecting such low peak velocity offset also implies negligible resonant scattering by residual neutral gas infalling onto the galaxy.
We also identify intense high ionization UV emission lines (C~{\small IV}, N~{\small IV}]) in this object, potentially indicating a hard radiation field, which could reduce the local residual neutral gas fraction.



\section*{Acknowledgments}

The authors acknowledge the anonymous referee for insightful comments which improved the manuscript. We thank Joris Witstok for useful discussions and Jorryt Matthee for kindly sharing data of the X-SHOOTER Lyman $\alpha$ survey at $z=2$ (XLS-$z2$; \citealt{Matthee2021}). We also thank St\'{e}phane Charlot and Jacopo Chevallard for providing access to the {\tt BEAGLE} tool used for SED fitting analysis. MT acknowledges funding from the {\it JWST} Arizona/Steward Postdoc in Early galaxies and Reionization (JASPER) Scholar contract at the University of Arizona. MT and RSE also acknowledge funding from the European Research Council under the European Union Horizon 2020 research and innovation program (grant agreement No. 669253). DPS acknowledges support from the National Science Foundation through the grant AST-2109066. CAM acknowledges support by the VILLUM FONDEN under grant 37459 and the Carlsberg Foundation under grant CF22-1322. The Cosmic Dawn Center (DAWN) is funded by the Danish National Research Foundation under grant DNRF140. ZL has been supported in part by grant AST-2009278 from the U.S. National Science Foundation.

The Binospec spectra used in this paper were obtained at the MMT Observatory, a joint facility of the University of Arizona and the Smithsonian Institution. We acknowledge the MMT queue observers for assisting with MMT/Binospec observations. The {\it HST} imaging and grism spectra used in this work are based on observations taken by the 3D-HST Treasury Program (GO 12177 and 12328) with the NASA/ESA {\it Hubble Space Telescope} obtained from the Space Telescope Science Institute (STScI), which is operated by the Association of Universities for Research in Astronomy, Inc., under NASA contract NAS 5-26555. This research is based in part on observations made with the NASA/ESA/CSA {\it James Webb Space Telescope} from the STScI under NASA contract NAS 5-03127. These observations are associated with programs \# 1180, 1181, 1210, 1345, and 3215. The {\it JWST} data were obtained from the Mikulski Archive for Space Telescopes at the STScI. 

%

\vspace{5mm}
\facilities{MMT (Binospec), {\it HST} (ACS and WFC3), {\it JWST} (NIRSpec)}


\software{{\tt numpy} \citep{Harris2020}, {\tt matplotlib} \citep{Hunter2007}, {\tt scipy} \citep{Virtanen2020}, {\tt astropy} \citep{AstropyCollaboration2013}, {\tt BEAGLE} \citep{Chevallard2016}, {\tt Cloudy} \citep{Ferland2013}}


\section*{Data Availability}

The {\it HST} and {\it JWST} data used in this work can be found in the Mikulski Archive for Space Telescopes (\url{https://mast.stsci.edu/}): \dataset[10.17909/T9JW9Z]{http://dx.doi.org/10.17909/T9JW9Z}, \dataset[10.17909/8tdj-8n28]{http://dx.doi.org/10.17909/8tdj-8n28}, and \dataset[10.17909/z7p0-8481]{http://dx.doi.org/10.17909/z7p0-8481}. Other data underlying this article will be shared on reasonable request to the corresponding author.





\bibliography{z23_resolved_Lya}{}
\bibliographystyle{aasjournal}



\end{document}